\documentclass[acmsmall]{acmart}

\usepackage{amsmath,amsfonts}

\usepackage{amssymb}
\usepackage{pdfpages}
\usepackage{textcomp}
\usepackage{xcolor}
\usepackage{xspace}
\usepackage{graphics}
\usepackage{epsfig}
\usepackage{enumitem}
\usepackage{booktabs}
\usepackage{multirow}
\usepackage{bbding}
\usepackage{url}
\usepackage{tikz}
\usepackage{filecontents}
\usepackage{pifont}
\usepackage{graphicx}
\usepackage{balance}  
\usepackage{subfig}
\usepackage{verbatim}
\usepackage{color,xcolor}
\usepackage{array}
\usepackage{bbding}
\usepackage{amsmath}
\usepackage{fancyhdr}

\usepackage{algorithm}
\usepackage[noend]{algpseudocode}
\usepackage{algpseudocode}

\newcommand\vldbdoi{XX.XX/XXX.XX}

\newcommand\vldbavailabilityurl{https://github.com/iDC-NEU/NeutronSparseLite}

\newcommand{\Paragraph} [1] {\smallskip\noindent{\bf #1 }}

\newcommand{\zyf}[1]{\textcolor{black}{#1}}
\newcommand{\wqg}[1]{\textcolor{black}{#1}}
\newcommand{\aix}[1]{\textcolor{black}{#1}}

\newcommand{\eat}[1]{}

\newcommand{\system}{NeutronSparse\xspace}

\begin{document}
\title{NeutronSparse: Coordinating Heterogeneous Engines for Sparse Matrix Multiplication on NPUs}

\author{Xin Ai}
\authornote{Both authors contributed equally to this research.}
\orcid{0009-0006-0746-8222}
\email{aixin0@stumail.neu.edu.cn}

\author{Zeyu Ling}
\authornotemark[1]
\orcid{0009-0006-9547-5755}
\email{lingzeyu@stumail.neu.edu.cn}

\author{Hao Yuan}
\orcid{0009-0002-6502-7696}
\email{yuanhao@stumail.neu.edu.cn}

\author{Qiange Wang}
\orcid{0000-0002-4847-6070}
\email{wangqiange94@gmail.com}
\affiliation{
  \institution{Northeastern University}
  \country{China}
}

\author{Yanfeng Zhang}
\orcid{0000-0002-9871-0304}
\authornote{Yanfeng Zhang is the corresponding author.}
\email{zhangyf@mail.neu.edu.cn}

\author{Yutao Peng}
\orcid{0009-0006-7999-5152}
\email{yujiayang@stumail.neu.edu.cn}

\author{Ge Yu}
\orcid{0000-0002-3171-8889}
\email{yuge@mail.neu.edu.cn}
\affiliation{
  \institution{Northeastern University}
  \country{China}
}

\begin{abstract}
Sparse matrix–matrix multiplication (SpMM) is a fundamental data operation for large-scale sparse data processing. With NPUs increasingly deployed in data centers for their performance and energy efficiency, accelerating SpMM on these platforms is a natural choice.
However, high-performance SpMM on NPUs poses a data management challenge, as irregular sparsity demands efficient data organization and scheduling. On Ascend 910B, the official MindSpore implementation achieves only 36.3\% of the performance of GPU-based sparse libraries such as cuSPARSE on NVIDIA A100.

To this end, we conduct an in-depth architectural analysis of SpMM execution on NPUs versus GPU and identify that the key performance bottleneck for SpMM on NPUs lies in the lack of efficient coordination across heterogeneous compute units under tile-based execution model. 
Therefore, we propose \system, a coordination-first SpMM framework for NPUs. 
\system integrates two key techniques: (i) Sparsity-aware coordination of heterogeneous engines, which adaptively partitions and balances workloads between heterogeneous compute units to keep them busy, and (ii) Locality-aware tile orchestrating, which reorganizes and reuses data tiles to reduce redundant computation and data movement overhead. 
Evaluations on Ascend 910B show that \system achieves 1.26$\times$-7.78$\times$ speedup over NPU baselines and 1.03$\times$–3.07$\times$ speedup over leading GPU libraries on 
NVIDIA A100, revealing untapped potential of NPUs for sparse computation.

\end{abstract}

\maketitle

\section{Introduction}

Sparse matrix–matrix multiplication (SpMM) is a fundamental kernel in modern data processing and data management systems \cite{nvidia2023cusparse,hongAdaptiveSparseTiling2019,yangDesignPrinciplesSparse2018,galeSparseGPUKernels2020,zhuSparseTensorCore2019,huangGESpMMGeneralPurposeSparse2020,zachariadisAcceleratingSparseMatrixMatrix2020,duffSurveySparseMatrix1977}, and is widely used in AI workloads \cite{sarkarFlowGNNDataflowArchitecture2022,yanRTGNNAcceleratingSparse2024,wangTCGNNBridgingSparse2021}. 
In particular, SpMM acts as a core data transformation that combines sparse structure with dense features, serving as the computational primitive for graph neural network aggregation \cite{neutron_sigmod_2022,hongtu_SIGMOD_2024,G3_SIGMOD_2023,SANCUS_VLDB_2022,NeutronOrch2024ai,aligraph_vldb_2019, neutrontp_vldb24} and sparse attention in large language models (LLMs) \cite{roy2021attention,kosurveyrecommendationsystems2022,waltersMolecularPropertyPrediction2020}. 
As input sizes grow from billion-scale graphs to LLMs with hundreds of billions of parameters \cite{yangDesignPrinciplesSparse2018,wangTCGNNBridgingSparse2021,yanRTGNNAcceleratingSparse2024}, SpMM increasingly stresses both compute and memory bandwidth, making accelerator support essential for scalable performance.

\begin{figure}[t]
  \centering
  \includegraphics[width=0.5\linewidth]{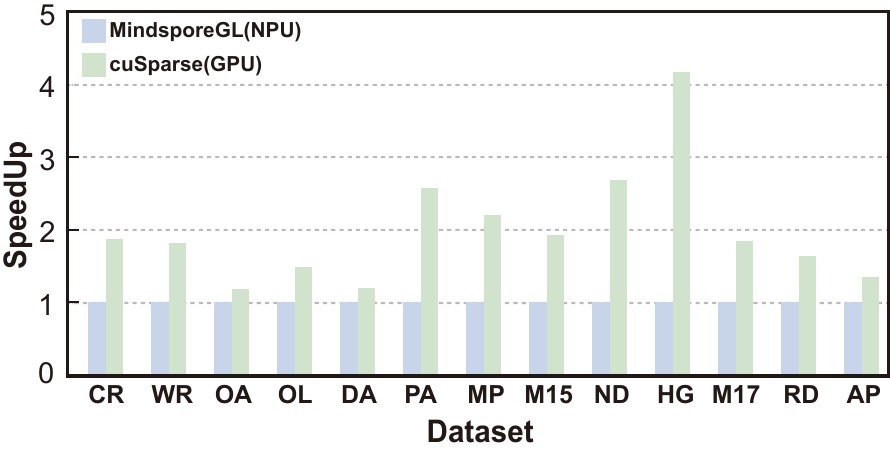}
  \vspace{-0.15in}
  \caption{Comparison of SpMM performance between the Ascend 910B and Nvidia A100.}
  \vspace{-0.2in}
  \label{GPUvsNPU}
\end{figure}

Neural Processing Units (NPUs) have become a cornerstone of modern AI infrastructure, with vendors such as Huawei (Ascend) \cite{Ascend_HPCA_2021}, Google (TPU) \cite{TPU_ISCA_2017}, Cambricon (MLU) \cite{MLU_MICRO_2021}, and Intel (Gaudi) \cite{Gaudi_Habana_2020} deploying their architectures widely across data centers.
As demand for AI computation accelerates, NPUs' market share continues to grow rapidly \cite{ascendwork1,ascendwork2}. 
These NPUs share two common architectural features. 
First, they are built around matrix-centric heterogeneous compute units (e.g., matrix, vector, and scalar engines) to support diverse operator and data-access patterns. 
Second, they follow a tile-centric execution style, where fixed-shape matrix/vector tiles are the basic unit of both computation and data staging.
These architectural features have the potential for higher computation capacity \cite{song2026xy}.
In this work, we focus on the Ascend NPU, which provides mature software support through its {Compute Architecture for Neural Networks (CANN)}.

Efficiently implementing SpMM on NPUs poses a data management challenge. Specifically, NPUs are architected around regular dataflows and high-throughput matrix engines \cite{heoNeuPIMsNPUPIMHeterogeneous2024,xiePatternBasedSpGEMMLibrary2022}, which perform well on dense tensors but struggle with the skew, dynamic sparsity, and irregular memory access that characterize SpMM workloads \cite{galeSparseGPUKernels2020,huangGESpMMGeneralPurposeSparse2020,sarkarFlowGNNDataflowArchitecture2022}. 
As shown in Figure \ref{GPUvsNPU}, our evaluation on the SuiteSparse dataset \cite{davis2011suitesparse} reveals that the official SpMM implementation in MindSpore \cite{zhuPerformanceEvaluationMindSpore2023} for Ascend 910B is 1.2 to 4.1 times slower than cuSPARSE running on an NVIDIA A100 GPU, despite both offering comparable peak FLOPS and memory bandwidth (910B: 280 TFLOPS, 1.8 TB/s; A100: 312 TFLOPS, 2.04 TB/s). 
\aix{This exposes the significant optimization gap in the current Ascend sparse computation stack and highlights the potential of architecture-aware data organization and execution orchestration for improving SpMM performance on sparse workloads.}
Given the widespread deployment of NPUs and the central role of SpMM in data-intensive AI pipelines (e.g., GNN aggregation and sparse attention), efficient SpMM on existing NPU platforms is essential to sustain scalable data management and analytics.

Unfortunately, SpMM optimization techniques developed for GPUs cannot be directly applied to NPUs due to fundamental architectural differences. 
We identify two NPU-specific challenges for SpMM. 
(1) heterogeneous engines must be explicitly coordinated to avoid load imbalance and idle compute. 
(2) tile-centric execution can introduce substantial redundant work and data movement overhead under irregular sparsity.

\begin{figure}[t]
  \centering
  \includegraphics[width=0.5\linewidth]{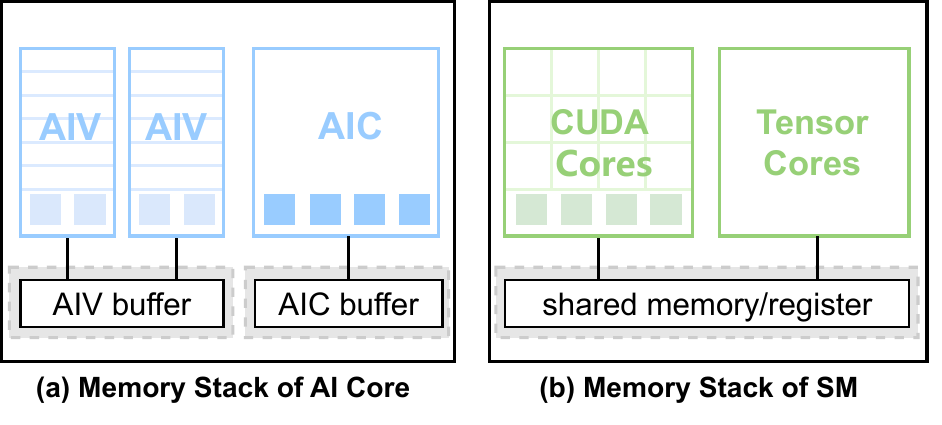}
    \vspace{-0.1in}
  \caption{Resource allocation in NPU AI Cores and GPU SMs.}
   \vspace{-0.2in}
  \label{diff_of_architecture}
\end{figure}

First, Ascend NPUs integrate a matrix engine (AIC) and a vector engine (AIV) within each AI Core.
Since they have independent instruction streams and per-engine local buffers, AIC and AIV can run concurrently with minimal on-chip resource contention (Figure~\ref{diff_of_architecture} (a))
\cite{Ascend_HPCA_2021,ascendwork1,ascendwork2,dharAscendCCConfidentialComputing2024}.
This makes coordination a first-order requirement for SpMM on NPUs: workloads must be partitioned and scheduled to keep AIC and AIV making balanced progress, otherwise one engine stalls.
By contrast, GPU SpMM optimizations typically alternates CUDA-core and Tensor-Core kernels with limited overlap, because both engines contend for shared SM resources (e.g., registers and shared memory) (Figure~\ref{diff_of_architecture} (b)) \cite{zhaoExploitingIntraSMParallelism2021}.
As a result, GPU approaches typically statically partition sparse and dense regions and execute the corresponding kernels sequentially
(e.g., sparse on CUDA cores and dense blocks on Tensor Cores)
\cite{wangTCGNNBridgingSparse2021,yanRTGNNAcceleratingSparse2024,liHCSpMMAcceleratingSparse2024,shi2025libra}.
Such mapping is fundamentally mismatched to NPUs: it routes work based on sparsity alone, ignoring heterogeneous engine throughput and execution balance. As a result, one engine dominates the critical path while the other idles, limiting effective overlap and parallelism

Second, Ascend NPUs follow a SIMD-based execution model that combines tile-based computation with software-managed memory \cite{Ascend_HPCA_2021}. AIC and AIV operate on fixed-size matrix and vector tiles, so SpMM is executed at a coarse, tile-level granularity. 
They always process tiles of a fixed shape, no matter how nonzeros are distributed inside the tile; when the sparsity pattern is uneven, many tiles contain only a few useful entries, which leads to redundant computation and load imbalance \cite{ahmadExploringDataLayout2024,shi2025flashsparse,merkelCanGraphReordering2024}.
GPUs instead follow a SIMT model, mapping thousands of lightweight threads to fine-grained work units (e.g., individual nonzeros or short segments) \cite{zhuSparseTensorCore2019,zhaoExploitingIntraSMParallelism2021}.  
SIMT-specific techniques such as warp-level NNZ load balancing (thread/warp-per-nonzero mapping) are hard to port to SIMD NPUs \cite{pangRoDe2024, shi2025libra, shi2025flashsparse, luoSparmSparseMatrix2024, zhaoExploitingIntraSMParallelism2021}.
Therefore, SpMM on NPUs calls for tile-centric scheduling that partitions and dispatches work at tile granularity to mitigate redundant work.

These differences introduce challenges for sparse workloads but also suggest an opportunity: SpMM on NPUs must be designed around coordinating heterogeneous engines under SIMD-based, tile-centric execution, rather than simply porting GPU-style kernels. 
Therefore, we propose \system, a coordination-first SpMM framework for NPUs.
First, it performs sparsity-aware coordination of heterogeneous engines that goes beyond data-driven partitioning. 
It derives a sparsity threshold based on hardware execution characteristics, which guides a lightweight row–column extraction that assigns sparse rows and columns to AIV and the remaining denser workloads to AIC.
At runtime, \system continuously monitors the relative progress of AIV and AIC and adaptively migrates tiles to the faster engine, ensuring that both engines make balanced progress and remain busy throughout the pipeline.
Second, \system introduces locality-aware tile orchestrating that is specifically designed for SIMD-style NPUs.
Rather than pursuing fine-grained NNZ-level reorderings, \system reshapes sparsity at tile granularity through a lightweight global-local reordering that forms denser tiles.
It further exploits software-visible control over the NPU memory hierarchy to enable hierarchical tile reusing, coordinating reuse across AI Cores via shared cache residency and within each AI Core via buffer-resident tile shaping.

Our contributions are summarized as follows:
\begin{itemize} [leftmargin=*]

  \item \textbf{Cross-Architecture Analysis.}
  We systematically compare SpMM on GPUs and Ascend NPUs, identifying two NPU-specific bottlenecks: cross-engine coordination and sparsity inefficiency under SIMD-based execution.

  \item \textbf{Sparsity-aware Heterogeneous Coordination.}
  We propose a coordination mechanism that combines architecture-aware static partitioning with runtime adaptive tile redistribution, balancing AIC/AIV workloads to maximize overlap.

  \item \textbf{Locality-aware Tile Orchestrating.}
  We propose a tile-granular data orchestrating method, combining global-local reordering with hierarchical tile reuse to reduce redundant computation and data movement overhead

  \item \textbf{The \system Framework.}
  We implement \system, an SpMM framework on Ascend 910B that integrates our coordination mechanisms with optimized AIC/AIV kernels.

\end{itemize}

We evaluate \system on an Ascend 910B processor. The experimental results show that \system achieves 1.26$\times$-7.78$\times$ speedup over the NPU baseline and achieves 1.03$\times$–3.07$\times$ speedup over state-of-the-art GPU SpMM libraries across various datasets.

The rest of this paper is organized as follows.
Section~\ref{sec2} introduces the architectures of GPUs and Ascend NPUs.
Section~\ref{sec3} presents our cross-architecture analysis and motivates NPU-specific SpMM design.
Section~\ref{sec4_1} describes sparsity-aware coordination of heterogeneous engines.
Section~\ref{sec5} describes our locality-aware tile orchestrating.
Section~\ref{sec6} describes the implementation of \system.
Section~\ref{sec7} evaluates \system on real-world workloads.
Section~\ref{sec8} discusses related work.
Section~\ref{sec9} concludes the paper.



\section{Background}
\label{sec2}

\subsection{GPU Architecture}
Modern GPUs are general-purpose parallel accelerators, commonly used in high-performance computing and AI workloads. 
They adopt the Single Instruction Multiple Thread (SIMT) programming model, where thousands of lightweight threads execute in parallel, grouped into warps of 32 threads that run in lock-step. 
Figure~\ref{GPU} shows the overview of the GPU architecture, using an NVIDIA GPU as an example.
At the hardware level, a GPU is composed of multiple Streaming Multiprocessors (SMs), each integrating scalar cores and, in some architectures, matrix-oriented cores such as NVIDIA Tensor Cores \cite{galeSparseGPUKernels2020,nvidia2023cusparse,hoImprovingGPUThroughput2022}. These compute units share the SM’s fast on-chip storage, including registers and shared memory \cite{galeSparseGPUKernels2020}. To further boost throughput, GPUs use hardware-managed L1/L2 caches and massive multi-threading to hide memory latency \cite{nvidia2023cublas,hoImprovingGPUThroughput2022,huangGESpMMGeneralPurposeSparse2020}. All SMs access a global memory space, enabling inter-SM communication and large-scale data processing \cite{zhaoAccSpMMAcceleratingGeneralpurpose2025}.

\subsection{Ascend NPU Architecture}
Ascend NPUs, developed by Huawei, are AI-specialized processors that adopt a heterogeneous design to accelerate deep learning and other compute-intensive tasks \cite{Ascend_HPCA_2021,dharAscendCCConfidentialComputing2024}.
Figure~\ref{AscendNPU} shows the overview of the Ascend NPU Architecture.
The core computational component is the AI Core, built on the DaVinci architecture \cite{DaVinci_HCS_2019, Ascend_HPCA_2021}, which integrates two primary compute engines: the AI Cube (AIC) and the AI Vector (AIV). AIC is optimized for dense matrix multiplications and features a 3D Cube Unit that processes 16×16×16 float16 tiles per cycle. AIV is optimized for vector operations and includes a 2048-bit Vector Unit capable of 128-wide computations per cycle \cite{Ascend_HPCA_2021}. Both engines are equipped with local buffers and scalar units for control and reduction operations. 
Each AI Core contains two AIV units and one AIC unit, each loading its own instruction stream to enable decoupled execution of vector and matrix tasks \cite{dharAscendCCConfidentialComputing2024,heoNeuPIMsNPUPIMHeterogeneous2024,Ascend_HPCA_2021,ascendwork1}. 
To facilitate data movement, Ascend provides a dedicated Memory Transfer Engine (MTE), which manages transfers between HBM and various on-chip buffers \cite{Ascend_HPCA_2021,ascendwork1}. 
In addition, Ascend includes a fixed-function FixPipe datapath that streams tiles between on-chip buffers and global memory for efficient staging and write-back (e.g., draining L0C results to HBM) \cite{Ascend_HPCA_2021,ascendwork1}.

\aix{Different generations of Ascend processors adopt slightly different AI Core organizations. Earlier Ascend architectures (Ascend 910A and early 910B) adopted a coupled design \cite{Ascend_HPCA_2021}, where the Cube unit can directly feed the Vector unit through Buffer C and the AI Core contains a single scalar control path. In contrast, this paper targets the newer decoupled design used by current mainstream Ascend processors (910B/910C/910D), where Cube and Vector are separated into AIC and AIV, the direct Cube-to-Vector path is removed, and each side has its own scalar unit for independent instruction dispatch and control. Figure~\ref{AscendNPU} follows this newer decoupled design.}

\section{Comparison and Analysis of GPU and NPU for SpMM}
\label{sec3}

While SpMM has been extensively optimized on GPUs \cite{hongAdaptiveSparseTiling2019,galeSparseGPUKernels2020,duffSurveySparseMatrix1977,zhuSparseTensorCore2019,luoSparmSparseMatrix2024,shi2025flashsparse,wangSwSpTRSVFastSparse2018,srivastavaMatRaptorSparseSparseMatrix2020,zachariadisAcceleratingSparseMatrixMatrix2020}, these optimizations do not directly carry over to NPUs due to key architectural differences.
We compare GPU and Ascend NPU architectures along two dimensions that fundamentally affect SpMM performance: \textbf{Compute Unit Heterogeneity} and \textbf{SIMD-based Execution}.




\subsection{Compute Unit Heterogeneity}

\begin{figure}[t]
  \centering

  \begin{minipage}[t]{0.48\linewidth}
    \centering
    \includegraphics[width=\linewidth]{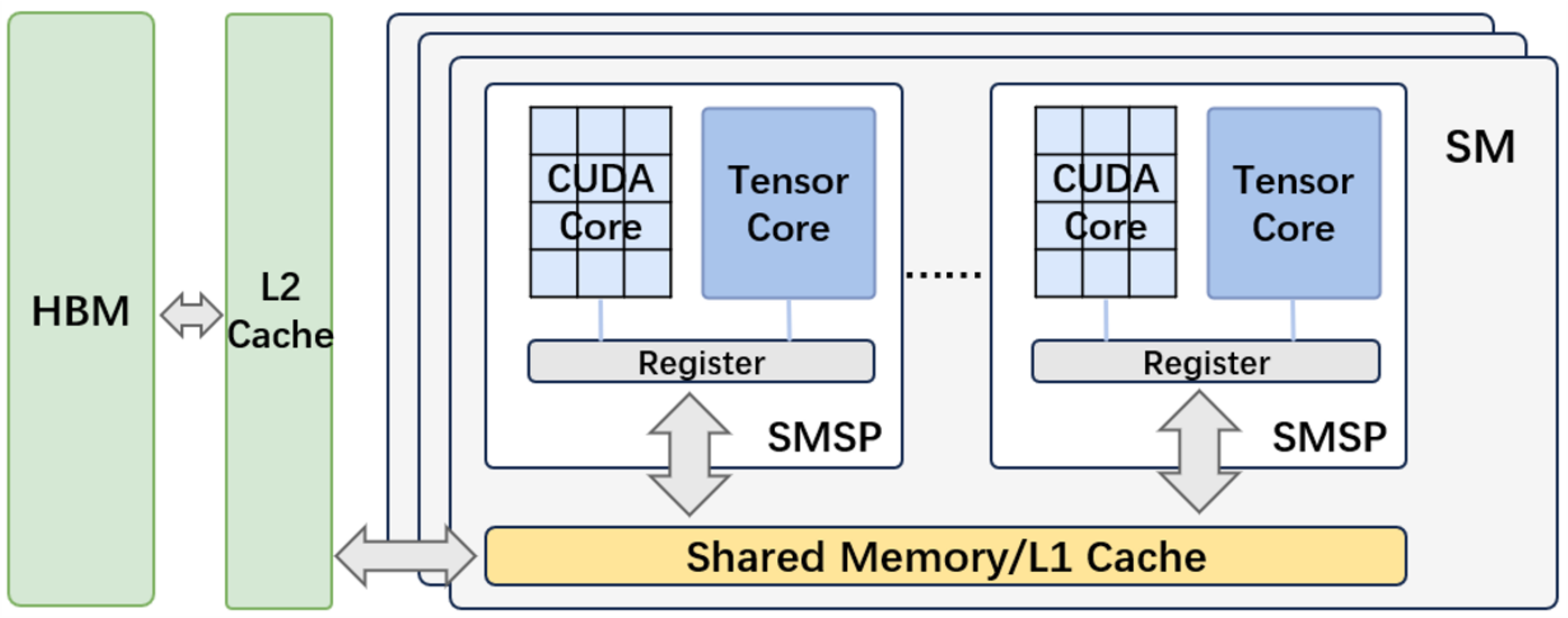}
    \caption{Overview of the Nvidia GPU architecture.}
    \label{GPU}
  \end{minipage}
  \hfill
  \begin{minipage}[t]{0.48\linewidth}
    \centering
    \includegraphics[width=\linewidth]{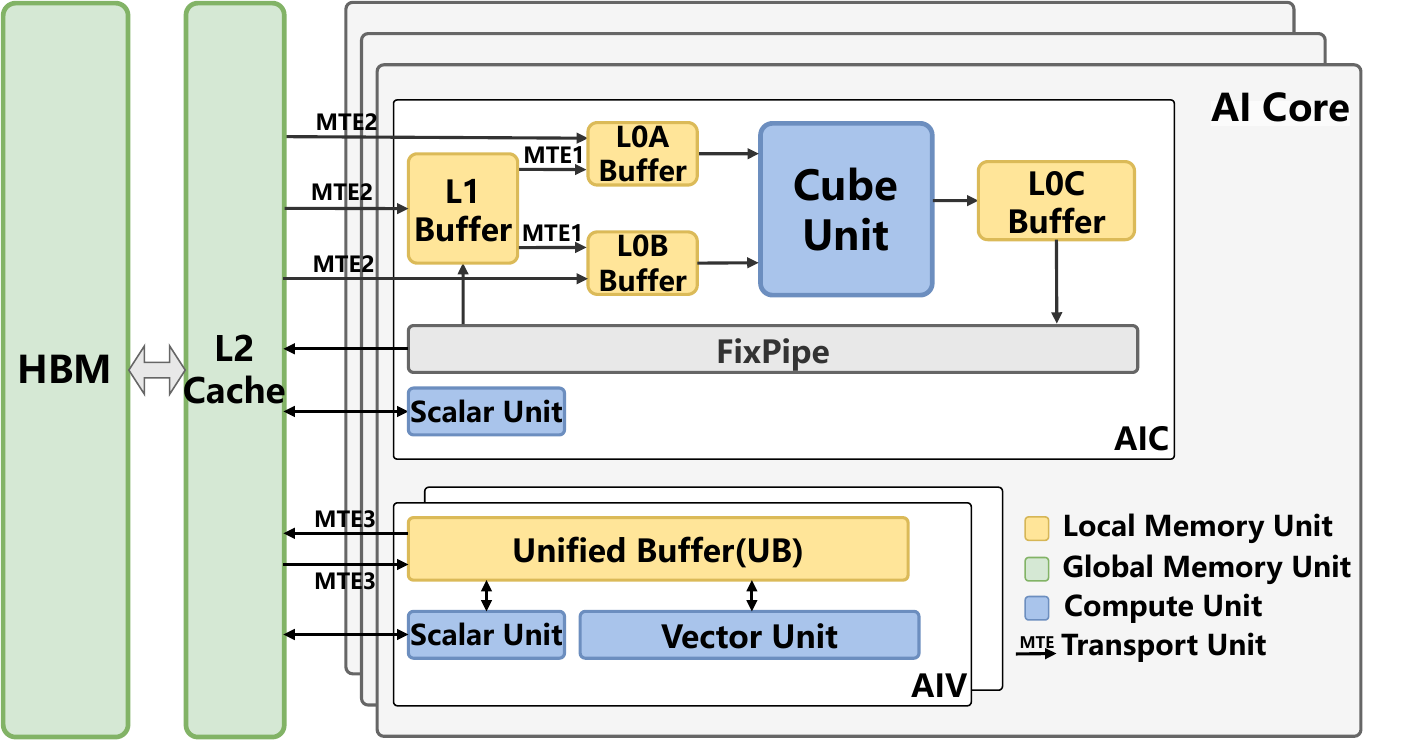}
    \caption{Overview of Ascend NPU architecture.}
    \label{AscendNPU}
  \end{minipage}

  \vspace{-0.1in}
\end{figure}

Both GPUs and Ascend NPUs employ heterogeneous compute units, but differ fundamentally in how these units interact and are scheduled. This distinction has a major impact on SpMM execution efficiency. Figure~\ref{GPU} and Figure~\ref{AscendNPU} highlight a key scheduling contrast between GPUs and Ascend NPUs.
On GPUs, CUDA cores and Tensor Cores are integrated within the same SM and contend for shared on-chip resources (e.g., registers/shared memory), which limits practical overlap in SpMM execution \cite{zhaoExploitingIntraSMParallelism2021, hoImprovingGPUThroughput2022}.
On Ascend, AIV and AIC are decoupled with independent pipelines and local buffers, enabling concurrent execution within an AI Core and making cross-engine coordination a primary lever for SpMM efficiency.


\begin{figure}[t]
  \centering
  \includegraphics[width=0.5\linewidth]{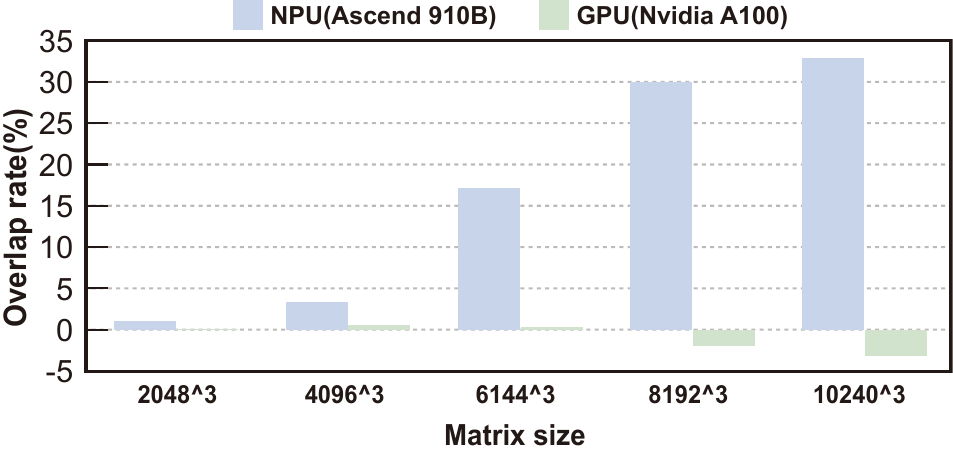}
  \vspace{-0.1in}
  \caption{Effective overlap between heterogeneous compute engines on NPU vs. GPU.}
  \label{OverlapRate}
  \vspace{-0.1in}  
\end{figure}

\begin{figure}[t]
  \centering
  \includegraphics[width=0.4\linewidth]{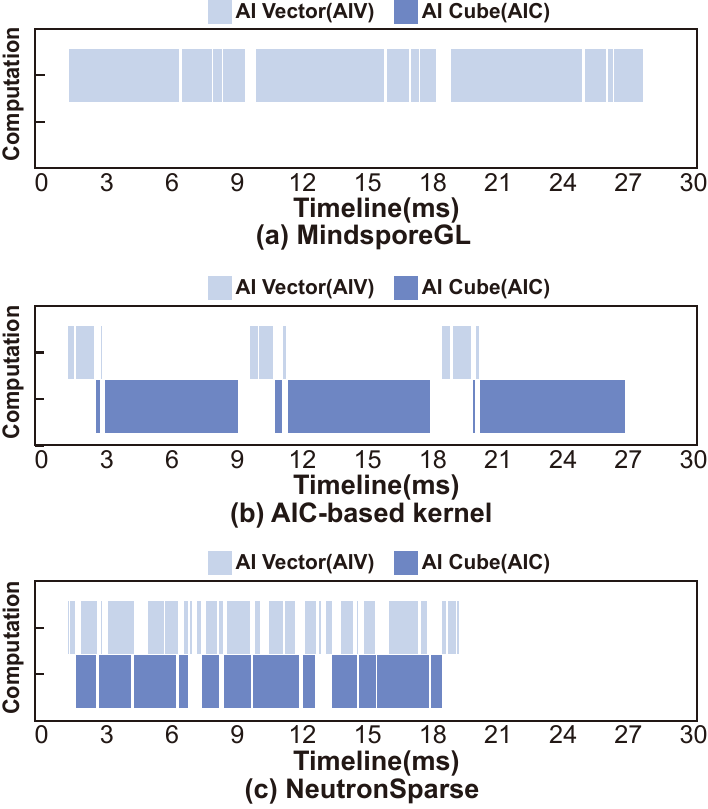}
  \vspace{-0.1in}
  \caption{\aix{AIC/AIV utilization timeline on \texttt{ogbn-arxiv}.}}
  \label{Utilization}
  \vspace{-0.1in}  
\end{figure}

\Paragraph{Experimental Analysis.} 
\aix{To quantify the benefit of concurrent execution across heterogeneous compute engines, we conduct a controlled overlap experiment on both NPUs and GPUs. We co-run two matrix multiplication tasks, assigning one to the vector-oriented engine (AIV/CUDA cores) and the other to the matrix engine (AIC/Tensor Cores), and vary the workload of the latter. We measure concurrency using the Overlap Rate metric \cite{zhaoExploitingIntraSMParallelism2021}, which captures the fraction of execution time saved due to overlap (upper-bounded by 50\%). As shown in Fig.~\ref{OverlapRate}, Ascend NPUs achieve consistently higher overlap (2.1\%-32.9\%), while GPUs show minimal or even negative overlap (-3.9\%-1.3\%). This gap indicates that NPUs provide substantially stronger support for cross-engine concurrency, motivating a coordination-first design for SpMM.}

\Paragraph{\aix{Limitation of Existing NPU Baselines.}}
\aix{We further analyze how prior Ascend SpMM implementations utilize heterogeneous engines within a fixed execution window on \texttt{ogbn-arxiv}. We profile the execution timeline of AIV and AIC and report their utilization patterns. As shown in Fig.~\ref{Utilization}, both baselines fail to exploit effective AIC–AIV concurrency. MindSporeGL executes almost entirely on AIV, with AIC utilization below 10\%, leaving the primary compute engine largely idle. In contrast, the AIC-based design heavily relies on AIC (over 80\% utilization), while AIV is only sporadically invoked, resulting in mostly serialized execution. In both cases, the overlap between AIC and AIV remains minimal, leading to underutilization of available compute resources. This analysis reveals a key limitation of existing designs: they rely on a single engine at a time rather than coordinating both engines. This further motivates a coordination-first approach that jointly schedules AIC and AIV to unlock the full potential of Ascend NPUs.}

\subsection{SIMD-based Execution}

GPUs follow Single Instruction Multiple Threads (SIMT), scheduling thousands of lightweight threads in warps to exploit fine-grained parallelism, whereas Ascend NPUs adopt Single Instruction Multiple Data (SIMD) and execute at a coarse, fixed-shape tile granularity through AIC/AIV \cite{zhaoExploitingIntraSMParallelism2021,galeSparseGPUKernels2020,Ascend_HPCA_2021}. 
This difference affects how SpMM handles irregular sparsity: SIMT can map work near the NNZ level, while SIMD issues whole tiles, so uneven nonzeros may translate into tile underutilization \cite{ahmadExploringDataLayout2024,shi2025flashsparse,merkelCanGraphReordering2024}.

Equally importantly for system design, SIMD-style NPUs expose substantially more software-visible control over data movement and residency in the on-chip hierarchy. 
To sustain SIMD throughput, tiles must be explicitly formed and staged across L2/L1/L0 buffers (e.g., via software-orchestrated transfers and cache controls), so reuse is not only a consequence of access patterns but also a controllable scheduling decision \cite{Ascend_HPCA_2021}. 
In contrast, GPUs rely largely on OS-managed caching and warp scheduling; locality is primarily improved by shaping thread-level access, while cache residency and eviction are mostly opaque to the programmer \cite{hoImprovingGPUThroughput2022,zhaoExploitingIntraSMParallelism2021}. 
Therefore, SpMM optimization on NPUs places greater emphasis on tile-centric organization and explicit buffer-aware scheduling to maximize reuse and reduce buffer traffic under the SIMD execution model \cite{ahmadExploringDataLayout2024,shi2025flashsparse,merkelCanGraphReordering2024}.

\begin{table}[t]
\centering
\caption{Fraction of redundant zeros within active tiles under different tile shapes. (TC: Tensor Core)}
\label{tab:redundancy}
\setlength{\tabcolsep}{4pt}
\small
\scriptsize
\begin{tabular}{l c c c c c}
\toprule
\multirow{2}{*}{Dataset} 
& \multicolumn{1}{c}{\textbf{GPU TC}} 
& \multicolumn{4}{c}{\textbf{NPU AIC Tiles}} \\
\cmidrule(lr){2-2} \cmidrule(lr){3-6}
& 4$\times$4 
& 16$\times$16 
& 32$\times$32 
& 64$\times$64 
& 128$\times$128 \\
\midrule
Cora        & 0.018 & 0.255 & 0.672 & 0.977 & 0.998 \\
Reddit      & 0.031 & 0.392 & 0.845 & 0.996 & 0.998 \\
Flickr      & 0.001 & 0.026 & 0.097 & 0.304 & 0.686 \\
Wiki-RfA    & 0.021 & 0.275 & 0.691 & 0.981 & 0.999 \\
Mouse\_gene & 0.153 & 0.724 & 0.929 & 0.986 & 0.996 \\
\midrule
\textbf{Average} 
& \textbf{0.04} 
& \textbf{0.33} 
& \textbf{0.64} 
& \textbf{0.84} 
& \textbf{0.92} \\
\bottomrule
\end{tabular}
\end{table}

\Paragraph{Experimental Analysis.}
To quantify the redundancy induced by tile-granular execution under sparsity, we conduct a controlled measurement on five real-world sparse matrices. We sweep the tile size $t \in \{4,16,32,64,128\}$ and conceptually tile each matrix; following a tile-centric execution abstraction, a tile is marked active if it contains at least one nonzero, and the kernel processes active tiles as whole $t{\times}t$ units. 
We then report the fraction of zeros inside all active tiles as a proxy for tile-induced redundant work. 
As summarized in Table \ref{tab:redundancy}, redundancy increases sharply with coarser tiles: averaged over the five datasets, the zero fraction rises from $0.04$ at $4{\times}4$ to $0.33$ ($16{\times}16$), $0.64$ ($32{\times}32$), $0.84$ ($64{\times}64$), and $0.92$ ($128{\times}128$). This trend highlights an architectural asymmetry: GPUs can exploit smaller tiles (e.g., $4{\times}4$ Tensor Core fragments) and fine-grained SIMT work mapping, keeping redundancy low and enabling flexible access to small blocks, whereas on Ascend NPUs the matrix engine operates at a much coarser granularity---$16{\times}16$ is the minimum AIC tile, making tile-level redundancy substantial even at the smallest supported tile size. Moreover, larger tiles are often favored on NPUs for better data reuse under SIMD-style execution, which further amplifies the redundancy penalty unless sparsity-aware tile organization and scheduling optimization are applied.

\begin{figure}[!t]
  \centering
  \includegraphics[width=0.75\linewidth]{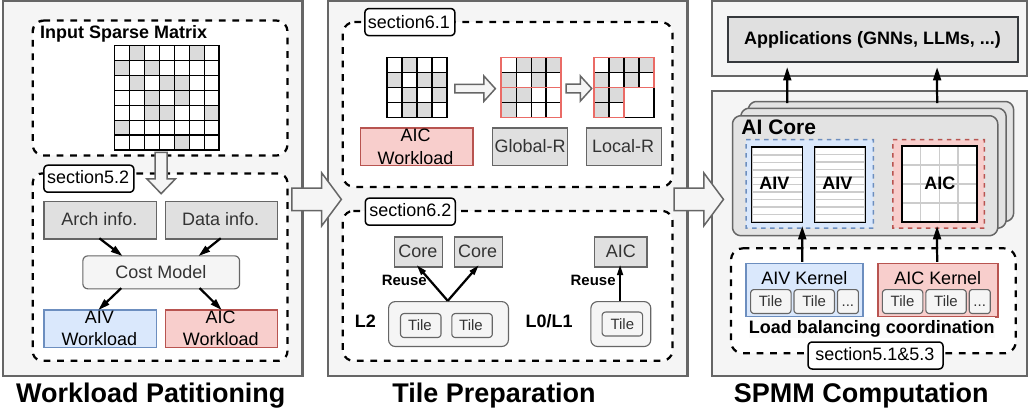}
  \caption{Workflow of \system.}
  \label{overview}
   \vspace{-0.1in}
\end{figure}

\section{The \system Framework}
\label{sec6}

\aix{We develop \system, an efficient SpMM framework on Ascend NPUs with two core functions: sparsity-aware coordination of heterogeneous engines and locality-aware tile orchestrating.}

\Paragraph{\aix{Overall Execution Flow.}} \aix{Figure~\ref{overview} summarizes the end-to-end workflow of \system, which turns an input sparse matrix into a coordinated AIC-AIV execution pipeline through three tightly coupled stages: workload partitioning, tile preparation, and SpMM computation. 
First, \system uses the cost model in Section 5.2 to split the SpMM workload into two engines, routing sparse tiles to AIV and dense tiles to AIC, thus establishing a balanced starting point on the heterogeneous engines. 
Next, it performs tile preparation to make this split executable under the NPU’s tile-based model: tiles are formed and optimized via global-local reordering (Section 6.1) to better align nonzeros at tile granularity, and tile reuse is explicitly planned across the memory hierarchy (Section 6.2) by staging frequently accessed dense $B$ rows for cross-core reuse while choosing per-core tile shapes that saturate private buffers and favor efficient transfers. 
Finally, AIV and AIC kernels (Section 5.1) consume the prepared tile streams concurrently within each AI Core, and \system applies the load-balancing coordination in Section 5.3 to overlap their progress and avoid engine idle time. 
}

\section{Sparsity-Aware Coordination of Heterogeneous Processing Engines}
\label{sec4_1}
We propose a sparsity-aware coordination strategy that combines static heterogeneous workload partitioning (Section~\ref{sec_partition}) with adaptive runtime coordinated pipelining (Section~\ref{sec_pipeline}), enabling sustained overlap between AIC and AIV.

\label{sec_execution_model}
\subsection{Execution Model of SpMM on AIC and AIV}
In this section, we establish a high-level execution model for SpMM on Ascend NPUs by characterizing how the heterogeneous AIC and AIV engines process sparse-dense multiplication. They follow fundamentally different execution models, reflecting their architectural specialization for dense and sparse workloads, respectively.

\begin{figure}[t]
  \centering
  \includegraphics[width=0.8\linewidth]{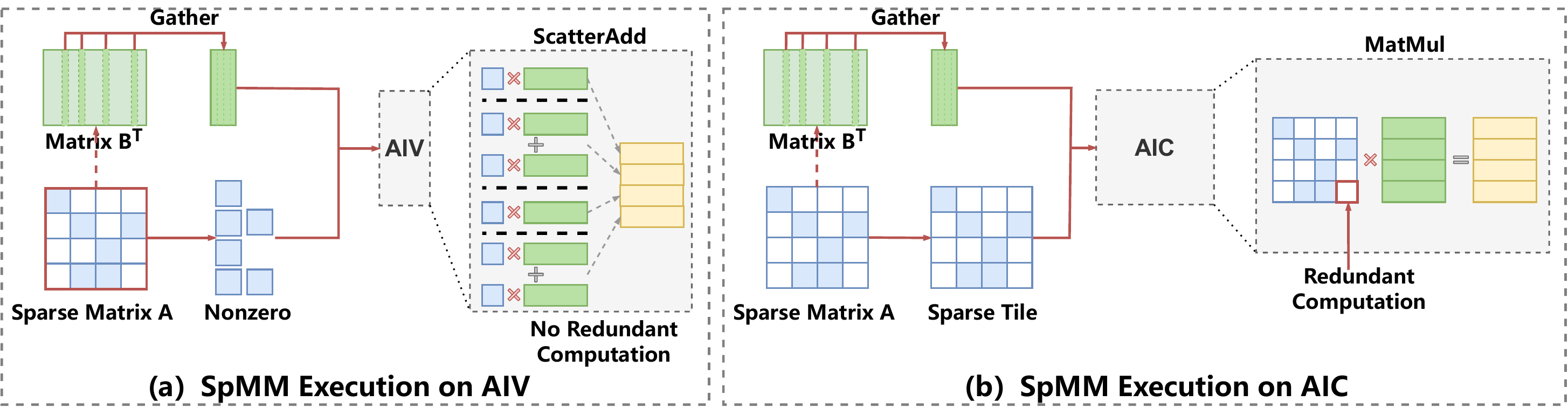}
  \vspace{-0.1in}
  \caption{
   \aix{SpMM execution on heterogeneous AIC and AIV engines: (a) Sparse, vector-oriented execution on AIV, (b) Dense, matrix-oriented execution on AIC.
  }}
   \vspace{-0.1in}
  \label{AIV_AIC_SpMM}
\end{figure}

\Paragraph{AIV-based SpMM Execution.}
As illustrated in Figure~\ref{AIV_AIC_SpMM}a, AIV executes SpMM in a vector-oriented manner. 
During execution, AIV performs a \textit{Gather} operation that uses the column indices of the sparse vector to fetch the corresponding row vectors from the dense matrix~$B$.
The gathered vectors are stored in the local on-chip buffer (UB) and subsequently accumulated into the output via \textit{ScatterAdd}. 
\aix{By operating directly on vectors after \textit{Gather}, AIV avoids redundant computation on zero entries and is therefore better suited to relatively sparse workloads.}

\Paragraph{AIC-based SpMM Execution.}
In contrast, AIC adopts a dense execution model optimized for matrix multiplication, as illustrated in Figure~\ref{AIV_AIC_SpMM}b.
During execution, the column indices associated with each tile of the sparse matrix~$A$ are used to perform a \textit{Gather} operation, which fetches the corresponding feature vectors from the dense matrix~$B$.
The gathered data are stored in the local on-chip buffer (L0B), while the corresponding tiles from matrix~$A$ are loaded into the local buffer (L0A).
\aix{Since computation is still performed on tile-level operands, zero entries inside sparse tiles also participate in \textit{MatMul}, introducing redundant work and making AIC more suitable for relatively dense tiles.}

\subsection{Heterogeneous Workload Partitioning}
\label{sec_partition}
The goal of heterogeneous workload partitioning is to decompose a sparse matrix into tile-level workloads and assign them to the processing engine best matched to their sparsity characteristics.
To this end, we first derive a sparsity threshold using an architecture-aware cost model, and then partition SpMM workloads across AIC and AIV accordingly.


\subsubsection{Architecture-Aware Cost Model}
\label{dry_run}

Existing partitioning methods for heterogeneous SpMM are largely data-centric, e.g., splitting workloads by sparsity and routing them to the preferred engine \cite{wangTCGNNBridgingSparse2021, yanRTGNNAcceleratingSparse2024, liHCSpMMAcceleratingSparse2024, shi2025libra}. On Ascend NPUs, this is insufficient because partitioning must optimize execution-time balance rather than data characteristics alone: even if AIV receives the sparse part and AIC receives the dense part, overall performance degrades whenever their assigned workloads finish at mismatched pace, leaving one engine idle and reducing overlap. Therefore, we introduce an architecture-aware cost model that estimates the sparsity tile runtime on AIV and AIC from hardware execution characteristics, and uses predicted runtime to guide workload allocation so that AIV and AIC make balanced progress during execution.

Specifically, we derive an architecture-aware sparsity threshold $\alpha$ via a lightweight sparse-aware dry run.
Here, $\alpha \in [0,1]$ denotes a density boundary (NNZ ratio) that separates tiles better handled by AIV from those better handled by AIC: tiles with density below $\alpha$ are assigned to the sparse-friendly AIV, while the remaining tiles are assigned to AIC.
The key is that this boundary should reflect real hardware efficiency, i.e., where AIV and AIC deliver comparable effective progress under their respective execution styles.

To obtain $\alpha$, we run SpMM microbenchmarks on both engines using representative strategies (AIC with GEMM-style dense multiplication and AIV with vectorized sparse-oriented multiplication), and measure their empirical throughputs, $P_{\text{AIC}}$ and $P_{\text{AIV}}$.
For a matrix of size $M \times K$ with $\text{NNZ}$ nonzeros, its density is $\text{NNZ}/(M\!\cdot\!K)$.
We model the execution time on AIV as proportional to the number of nonzeros, and the execution time on AIC as proportional to the full tile volume:
\begin{equation}
\text{Cost}_{\text{AIV}}(\text{NNZ}) = \frac{\text{NNZ}}{P_{\text{AIV}}}, \quad
\text{Cost}_{\text{AIC}} = \frac{M \cdot K}{P_{\text{AIC}}}.
\end{equation}

We then choose the crossover point $\text{NNZ}^*$ such that the two engines achieve a \emph{balanced progress rate} according to their architectural capacity ratio $r$ (e.g., $r{=}2$ since each AI Core contains two AIV units and one AIC unit).
Concretely, we find $\text{NNZ}^*$ that makes $\text{Cost}_{\text{AIV}}(\text{NNZ})/\text{Cost}_{\text{AIC}}$ as close as possible to $r$:
\begin{equation}
\text{NNZ}^* = \arg\min_{\text{NNZ}} \left(\frac{\text{Cost}_{\text{AIV}}(\text{NNZ})}{\text{Cost}_{\text{AIC}}} - r \right)^2.
\end{equation}
Minimizing this objective yields
$
\frac{\text{NNZ}^*}{P_{\text{AIV}}} = r \cdot \frac{M \cdot K}{P_{\text{AIC}}},
$
and thus the sparsity threshold
\begin{equation}
\alpha = \frac{\text{NNZ}^*}{M \cdot K} = r \cdot \frac{P_{\text{AIV}}}{P_{\text{AIC}}}.
\end{equation}
In the following, we use $\alpha$ to classify tiles by density and construct tile-level workloads for heterogeneous execution.

\begin{figure}[!t]
  \centering
  \includegraphics[width=0.6\linewidth]{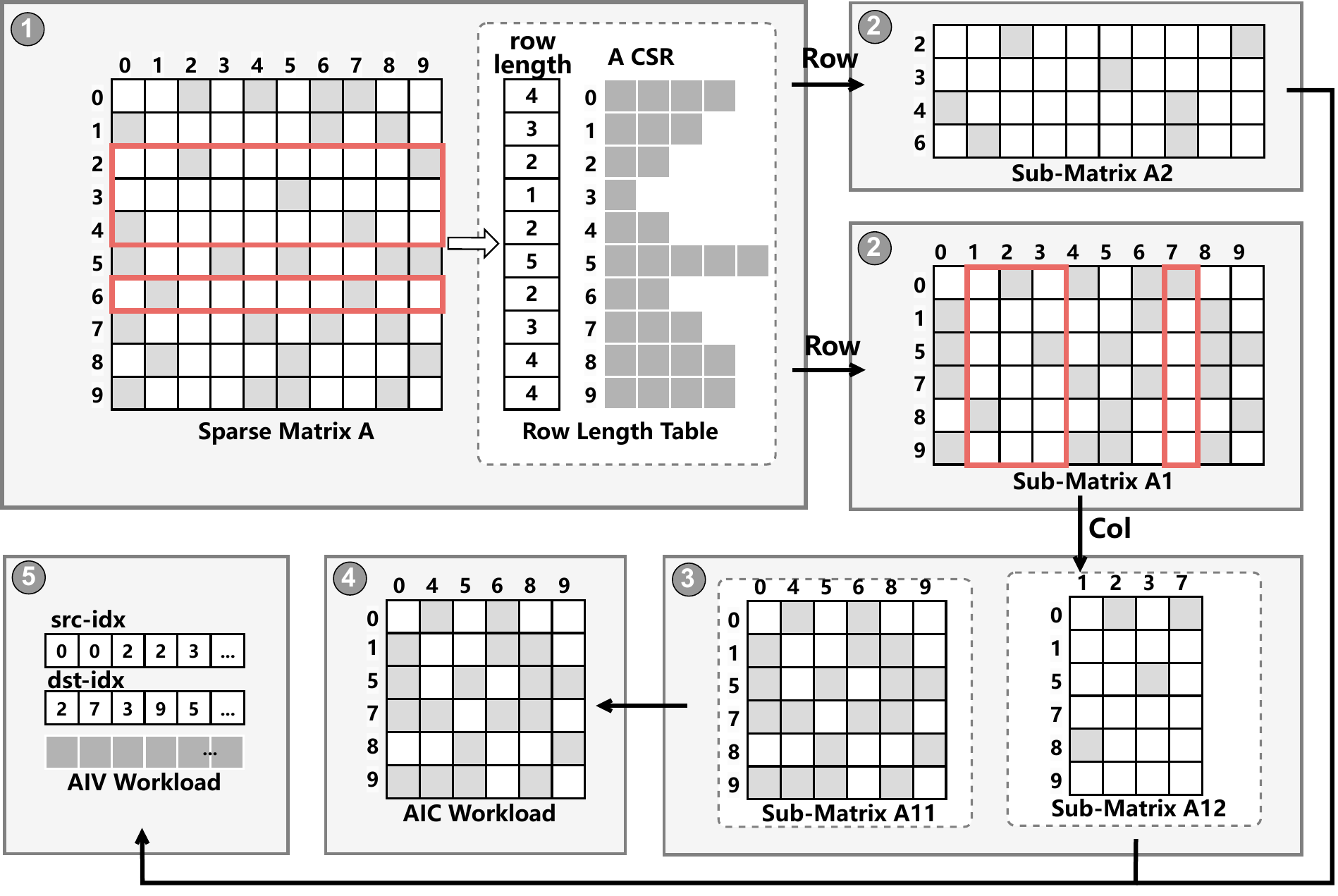}
  \vspace{-0.1in}
  \caption{Heterogeneous workload partitioning.
  The sparse matrix is first partitioned by rows into a dense part ($A_1$) and a sparse part ($A_2$), and then $A_1$ is further partitioned by columns into $A_{11}$ and $A_{12}$.
  }
  \label{TaskPartition}
  \vspace{-0.1in}
\end{figure}

\subsubsection{AIV-AIC Workload Partitioning}

Using the density threshold $\alpha$, we partition SpMM workloads across AIV and AIC via a lightweight two-stage {row--column extraction}.
We choose extraction-based partitioning because it (i) can be implemented by a single linear scan over the sparse structure, (ii) directly targets workload skew dominated by a few long rows/columns, and (iii) yields submatrices that naturally match the data formats and execution paths of AIV (irregular sparse vectors) and AIC (regularized dense tiles).

\Paragraph{Extraction Metric and Rule.}
Let $v$ denote a sparse {vector} in $A$ (a row or a column, depending on the extraction dimension).
We define its nonzero length as
\begin{equation}
\mathit{Len}(v) = \left|\{\, x_j \mid x_j \in v,\; x_j \neq 0 \}\right|,
\label{eq:Len}
\end{equation}
and convert the density boundary $\alpha$ into a length threshold
\begin{equation}
\mathit{Thres} = \alpha \cdot K,
\label{eq:Thres}
\end{equation}
where $K$ is the vector dimension.
Vectors with $\mathit{Len}(v) \le \mathit{Thres}$ are treated as {sparse workloads} and assigned to AIV, while the remaining vectors are assigned to AIC.

\Paragraph{Two-Stage Row-Column Extraction.}
As shown in Figure \ref{TaskPartition}, we apply the above rule in two stages.
We first traverse $A$ once to compute $\mathit{Len}(\cdot)$ and extract sparse {rows} to AIV; the remaining rows form a denser submatrix are routed to AIC.
Since row extraction alone may still leave the AIC submatrix with highly sparse {columns}, we apply the same extraction rule once more along the column dimension {within the AIC submatrix}, moving short columns back to AIV.
This two-stage procedure decomposes $A$ into a dense core handled by AIC and sparse fringes handled by AIV, providing a static, sparsity-aware workload decomposition that is well aligned with heterogeneous execution.
We store the AIC-assigned dense core in a BitMap to enable sequential access and tile-based computation, while storing the AIV-assigned sparse fringes in COO to avoid zero storage and support irregular gather/scatter accesses.

\subsection{Adaptive AIV-AIC Coordinated Pipelining}
\label{sec_pipeline}
Static partitioning (Section~\ref{sec_partition}) provides an initial AIV/AIC mapping guided by the hardware-aware threshold $\alpha$.
During execution, however, SpMM on Ascend runs as a concurrent two-engine pipeline: AIV and AIC process their assigned workloads in parallel within an AI Core, and the end-to-end latency of each iteration is determined by the slower engine.
Due to irregular sparsity and runtime variation (e.g., skewed rows/columns and cache effects), the two engines may drift out of balance, causing the faster engine to stall and leaving pipeline bubbles.
We therefore introduce an adaptive coordinated pipeline that (i) monitors the relative progress of AIV and AIC at coarse granularity and (ii) migrates a small amount of {remaining work} from the bottleneck engine to the idle engine, thereby sustaining overlap and minimizing idle gaps.
\begin{figure}[t]
  \centering
  \includegraphics[width=0.6\linewidth]{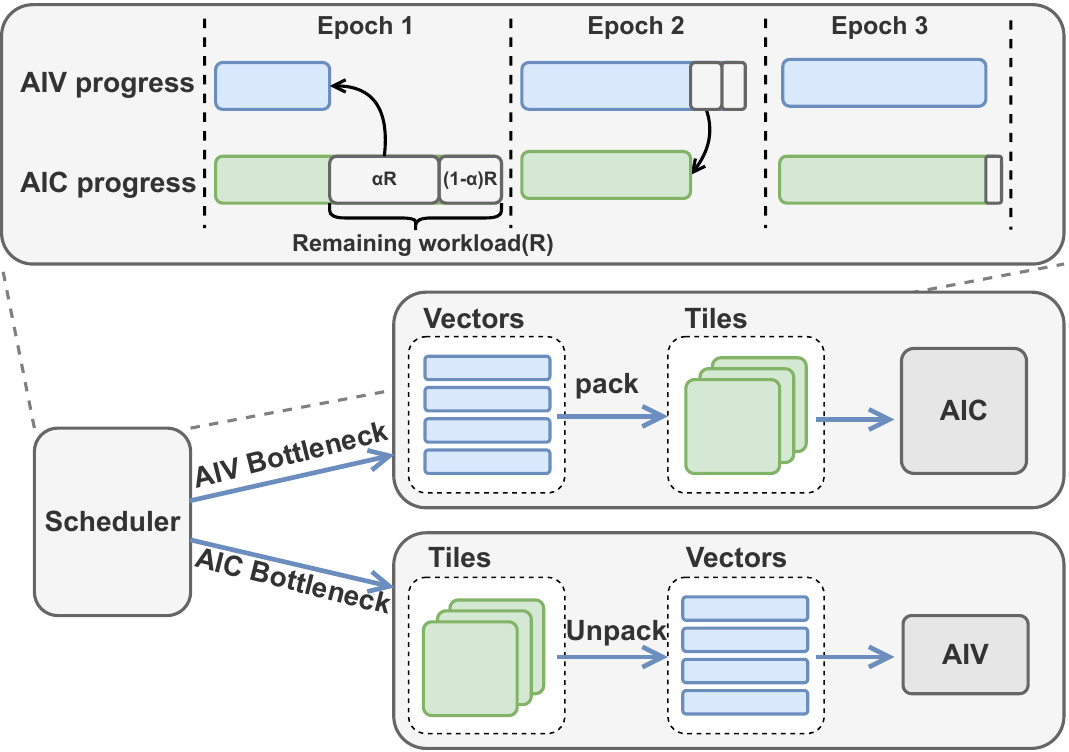}
  \vspace{-0.1in}
  \caption{Load migration under runtime load imbalance.
   }
  \label{Migration}
  \vspace{-0.1in}
\end{figure}

\Paragraph{Progress Monitoring at Epoch Granularity.}
Many target workloads repeatedly invoke SpMM across a large number of iterations/epochs (e.g., GNN training typically runs hundreds of epochs and sparse attention executes similar sparsity patterns across layers and steps).
This repetition makes epoch-level adjustment effective: a decision made from the previous epoch's timing feedback remains informative for subsequent epochs, while keeping the monitoring overhead negligible.
Accordingly, we execute SpMM in epochs and collect per-epoch execution time for each engine:
$\Delta t_{\text{AIC}}$ for AIC and $\Delta t_{\text{AIV}}$ for AIV (Figure~\ref{Migration}, top).
We quantify imbalance with a dimensionless skew ratio
\begin{equation}
\label{eq:skew}
\text{Skew} \triangleq 
\frac{\max(\Delta t_{\text{AIC}},\, \Delta t_{\text{AIV}})}
{\min(\Delta t_{\text{AIC}},\, \Delta t_{\text{AIV}})} \;\; \ge 1 .
\end{equation}
When $\text{Skew} \le 1+\epsilon$, the engines are considered sufficiently balanced.
\aix{We trigger migration only if $\text{Skew} > 1+\epsilon$, where $\epsilon$ is a small tolerance (e.g., 0.05) to avoid oscillation.
This lightweight mechanism serves as a short warm-up process after the initial partition, using only epoch-level timing to correct residual imbalance with negligible overhead. 
Our current target is mainly static SpMM workloads. For dynamic workloads, matrix evolution is often gradual, so the original partition can remain effective for multiple epochs and migration can be re-triggered after sufficient changes accumulate.}

\Paragraph{Adaptive Workload Migration.}
When imbalance is detected, we migrate part of the {residual workload} from the bottleneck engine to the other engine to eliminate pipeline bubbles.
Let $R$ denote the remaining work on the slower engine at the moment the faster engine would otherwise become idle (Figure~\ref{Migration}, top).
Migration follows a {sparsity-guided direction} consistent with the execution models of AIV and AIC (Figure~\ref{Migration}, bottom):

\begin{itemize}[leftmargin=1.2em]
\item \textbf{AIC is the bottleneck ($\Delta t_{\text{AIC}} > \Delta t_{\text{AIV}}$).}
We preferentially {extract sparser tiles} from AIC's residual work and offload them to AIV.
Concretely, we decompose candidate tiles into sparse vectors (index-value lists).

\item \textbf{AIV is the bottleneck ($\Delta t_{\text{AIV}} > \Delta t_{\text{AIC}}$).}
We preferentially {densify and pack relatively dense sparse vectors} from AIV's residual work into tile-aligned matrix blocks and offload them to AIC.
Concretely, we merge denser rows/segments to form matrix tiles.
\end{itemize}

\noindent \zyf{During the local reordering stage (Sec.~6.1), when the matrix is organized into fixed-shape tiles, we simultaneously record the sparsity of each tile. Online migration then directly uses these precomputed sparsity values to identify sparser tiles from the residual workload.}

We use the threshold $\alpha$ (Section~\ref{dry_run}) as a hardware-aware target split between AIV- and AIC-suited workloads. Given the residual workload $R$, we partition it into $R'_{\text{AIV}}$ and $R'_{\text{AIC}}$ to balance execution time across the two engines.
Since AIV and AIC follow different cost models (Eq.~(1)), the partition is not based on row or tile counts. Instead, we select $R'_{\text{AIV}} \subseteq R$ such that its nonzero count and the remaining dense workload assigned to AIC satisfy:
\begin{equation}
\frac{\text{NNZ}(R'_{\text{AIV}})}{M(R'_{\text{AIC}}) \cdot K} 
\;\approx\; 
\alpha \;=\; r \cdot \frac{P_{\text{AIV}}}{P_{\text{AIC}}}.
\end{equation}
where $R'_{\text{AIC}} = R \setminus R'_{\text{AIV}}$, and $M(R'_{\text{AIC}})$ denotes the number of rows in the AIC-assigned workload. 
This condition aligns the workloads according to the throughput capabilities of the two engines and corresponds to the sparsity threshold $\alpha$.

\section{Locality-Aware Tile Orchestrating}
\label{sec5}

On NPUs, SpMM is fundamentally governed by the tile as the minimum unit of computation and data movement.
While many GPU-style reorderings can in principle be applied to NPUs, directly reusing NNZ-/thread-level transformations is often unnecessarily fine-grained for tile execution and can introduce prohibitive preprocessing overhead.
This motivates a different optimization principle for NPUs: improve locality at tile granularity, and trade heavy preprocessing for lightweight, hardware-effective transformations.
Equally importantly, SIMD-style NPUs expose much more explicit control over tile movement and residency across on-chip buffers.
Since tile materialization, staging, and eviction are largely software-orchestrated, performance depends not only on how tiles are formed, but also on where they are kept and how often they are reused under limited buffer capacity.

Therefore, we adopt a lightweight, tile-centric orchestrating strategy that jointly optimizes tile formation and tile reuse to reduce both redundant tile execution and unnecessary memory traffic.
Specifically, we introduce two complementary mechanisms: (i) a global-local tile reordering scheme that coarsely reshapes row/column locality and then cheaply refines tile alignment to form denser, better-aligned tiles, and (ii) {hierarchical tile reusing} that improves reuse both {across AI Cores} (by retaining high-frequency dense $B$ tiles in shared cache) and {within each AI Core} (by choosing tile shapes that maximize buffer residency and enabling aligned write-back).
Together, these techniques improve sustained throughput under irregular sparsity.

\label{sec_reorder}

\begin{figure}[t]
  \centering
  \includegraphics[width=0.6\linewidth]{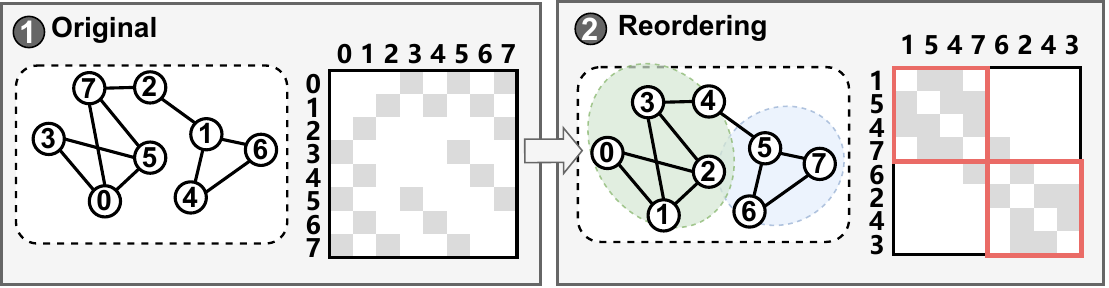}
  \vspace{-0.1in}
  \caption{Global Reordering.
  }
   \vspace{-0.2in}
  \label{global_reorder}

\end{figure}

\subsection{Global-Local Tile Reordering}
\label{sec:global_local_reordering}



Prior reorderings for GPUs often rely on fine-grained row/column permutations to maximize NNZ-level alignment and clustering~\cite{merkelCanGraphReordering2024,liHCSpMMAcceleratingSparse2024,zhaoAccSpMMAcceleratingGeneralpurpose2025,fanDTCSpMMBridgingGap2024}. 
For tile-based NPUs, we instead target {tile-level} density and alignment with a lightweight two-level {global-local} reordering scheme.
The \textbf{global stage} performs a coarse permutation over both rows and columns to form a small number of large clusters, capturing dominant structural correlations with low overhead.
The \textbf{local stage} then refines each cluster at {row-window} granularity: it groups rows with similar sparsity patterns to improve intra-tile alignment and to facilitate column compaction during tile construction.
Together, the two stages achieve tile-friendly layouts while keeping preprocessing lightweight.


\Paragraph{Global Reordering.}
\wqg{The global stage aims to expose coarse structural locality by grouping strongly related rows and columns into clusters. 
Here, rows are considered related if they share similar nonzero-column patterns, while columns are considered related if they share similar nonzero-row patterns.
As shown in Figure \ref{global_reorder}, we model the sparse matrix as a bipartite graph and apply Rabbit Order~\cite{merkelCanGraphReordering2024,arai2016rabbit} to reorder the original row and column indices so that related rows and columns become adjacent and form clusters. Unlike aggressive clustering schemes that iterate until convergence \cite{merkelCanGraphReordering2024,liHCSpMMAcceleratingSparse2024,zhaoAccSpMMAcceleratingGeneralpurpose2025}, we intentionally limit the number of clusters.
This reordering makes originally scattered nonzeros fall into more localized row and column ranges, forming denser block-like regions in the reordered matrix. Such a coarse permutation captures the major structural correlations needed by the subsequent local reordering, while keeping preprocessing overhead low and avoiding unnecessary fragmentation for tile-oriented execution.}

\begin{figure}[t]
  \centering
  \includegraphics[width=0.6\linewidth]{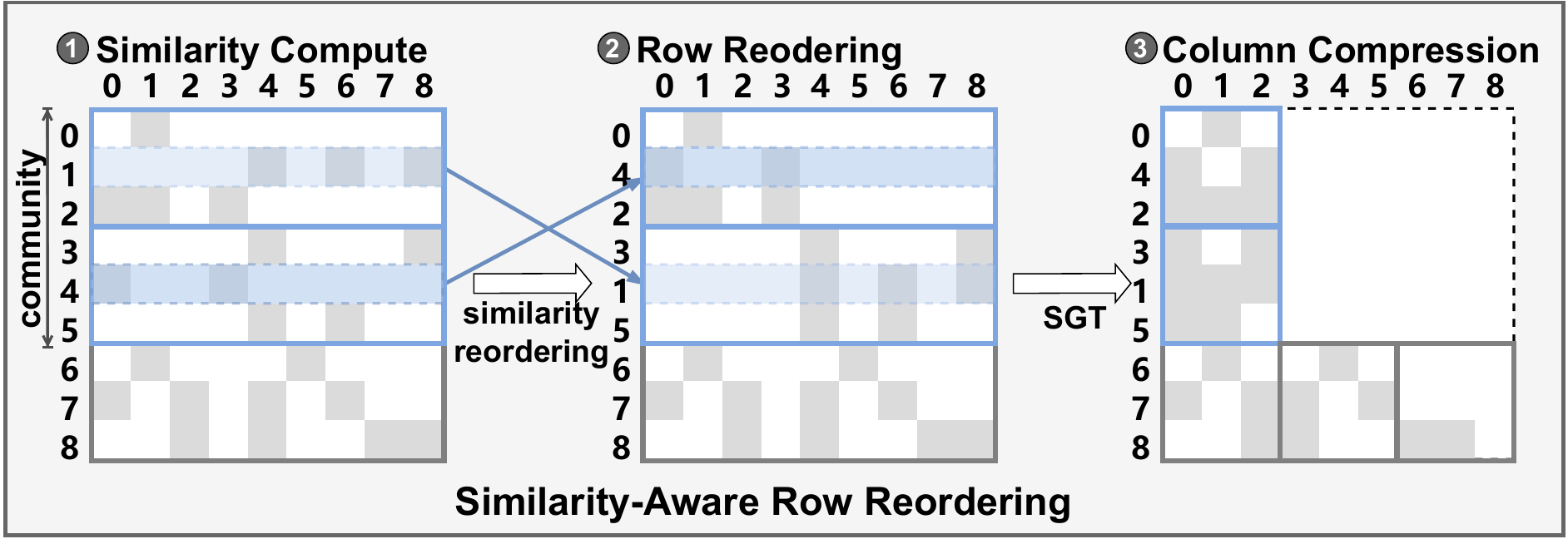}
   \vspace{-0.1in}
\caption{Local Reordering.
}
  \label{local_reorder}
  \vspace{-0.2in}
\end{figure}

\Paragraph{Local Reordering.}
After global clustering, residual sparsity still exists inside each cluster. Rather than continuing expensive element-level permutations over both rows and columns, the local stage refines locality at tile granularity. As shown in Figure~\ref{local_reorder}, we partition each cluster into row windows whose height matches the target tile shape (i.e., $M$, set to 128 in Section~5.2). We then reorder rows \emph{within each cluster} by grouping rows with similar sparsity patterns into the same window, which increases nonzero alignment across columns and exposes empty column regions that can be compacted during tile construction. Concretely, we use the current row windows as anchors by selecting one representative row per window, and then rebuild windows greedily: for each anchor row, we pick the $(M{-}1)$ most similar unassigned rows based on Jaccard similarity over their nonzero column sets, form a new window, and remove assigned rows to avoid duplication. Importantly, this refinement only permutes rows and does not alter the global column order, making it much cheaper than full element-level reordering while still producing denser, better-aligned tiles.

In summary, the global--local reordering strategy reflects a deliberate trade-off.
The global stage provides a coarse approximation of element-level reordering to capture long-range structure, while the local stage performs low-cost, tile-oriented refinement to align with SIMD execution.
This design contrasts with GPU-oriented approaches that pursue maximal fine-grained alignment, and instead tailors reordering to the needs and limits of tile-based NPUs.

\subsection{Hierarchical Tile Reusing}
\label{sec_tile_reuse}

Tile reusing on NPUs can be reasoned at two distinct levels: reusing across AI Cores and reusing within a single AI Core.

\subsubsection{Inter-core Tile Reusing.}
After partitioning the sparse matrix $A$ by rows (or row windows), different AI Cores process disjoint subsets of $A$.
However, both AIV and AIC repeatedly perform \texttt{Gather} to fetch rows of the dense matrix $B$ according to the column indices of nonzeros in $A$.
As a result, even when AI Cores work on independent tiles of $A$, their accesses to $B$ exhibit substantial overlap.
This makes $B$ a natural target for inter-core tile reusing.

Unlike GPUs, where L2 cache behavior is largely managed by hardware and the operating system, Ascend NPUs allow software to explicitly control whether memory accesses are cached in L2 \cite{Ascend_HPCA_2021,AIXIN2025neutronascend}.
We exploit this capability to orchestrate $B$ as a shared working set across AI Cores.
Guided by the global reordering from Section 5.1, we process the tiles cluster by cluster.
For each active cluster, we identify the rows of $B$ that are most frequently accessed by tiles within that cluster and proactively stage them into the shared L2 cache.
During the processing of this cluster, accesses to these high-frequency rows of $B$ are marked as cacheable, while accesses to sparse tiles of $A$ and less frequently used rows of $B$ bypass L2.

The size of the staged $B$ working set is bounded by the L2 capacity (typically 20--32\,MB).
In practice, we conservatively cap it at approximately 80\% of the available L2 space to tolerate system-reserved entries and runtime fluctuations.
This design enables multiple AI Cores to reuse the same $B$ tiles while processing different row windows, improving the reuse efficiency of hot $B$ tiles and reducing redundant HBM traffic induced by repeated \texttt{Gather}.

\subsubsection{Intra-core Tile Reusing.}
Within an AI Core, tile reusing mainly takes the form of {tile shaping}, where the tile shape determines how much computation can be performed before operands and partial results are evicted from on-chip buffers.
By keeping data resident longer, a well-orchestrated tile improves intra-core reuse efficiency.
In our design, each tile corresponds to one row window produced by local reordering, with height $M$ and width $K$ in the sparse window, and width $N$ in the associated dense matrix $B$.
Selecting a proper $(M,N,K)$ requires balancing compute throughput, buffer capacity limits, and write-back efficiency.

\Paragraph{Architectural Constraints.}
Ascend~910B provides three on-chip buffers per AIC: L0A and L0B (64\,KB each) for fp16 operands of the $M{\times}K$ $A$-tile and the $K{\times}N$ $B$-tile, and L0C (256\,KB) for fp32 accumulation of the $M{\times}N$ output tile.
To overlap computation with data transfer, we employ double-buffer pipelining (see Section~6.2 for details), so a tile can use at most half of each buffer.
Let $C_{A}{=}C_{B}{=}64$\,KB and $C_{C}{=}256$\,KB denote the capacities of L0A, L0B, and L0C, respectively.
Since fp16 and fp32 occupy $2$ and $4$ bytes per element, the residency constraints are:
\begin{equation}
\label{eq:l0_constraints}
\underbrace{(M K)\cdot 2}_{A\text{-tile bytes}} \le \tfrac{1}{2}C_{A},\qquad
\underbrace{(N K)\cdot 2}_{B\text{-tile bytes}} \le \tfrac{1}{2}C_{B},\qquad
\underbrace{(M N)\cdot 4}_{C\text{-tile bytes}} \le \tfrac{1}{2}C_{C}.
\end{equation}
Substituting $C_{A}{=}C_{B}{=}64$\,KB and $C_{C}{=}256$\,KB yields:
\[
MK \le 16384,\qquad NK \le 16384,\qquad MN \le 32768,
\]
where $M$, $N$, and $K$ must also be multiples of 16 to satisfy AIC execution requirements.

\Paragraph{Write-back Preference.}
The output matrix is stored in row-major layout and written back row by row.
As a result, consecutive output tiles advance the global write pointer by $N \!\cdot\! 4$ bytes.
When this stride is aligned to the fixpipe 512-byte transaction granularity, write-back achieves higher bandwidth; otherwise, split transactions may degrade efficiency~\cite{Ascend_HPCA_2021,ascendwork1}.
This favors tile shapes with $N$ being a multiple of 128.

\Paragraph{Maximizing AIC Utilization.}
AIC throughput is determined by the tile volume $MNK$.
Given the input-buffer constraints $MK \le 16384$ and $NK \le 16384$, the maximum achievable $MNK$ is reached when both bounds are tight.
One such configuration is $(128,128,128)$, which fully utilizes L0A and L0B and satisfies the preferred $N{\ge}128$ for efficient write-back.

From this peak-throughput point, we can increase $N$ and reduce $K$ proportionally while keeping $NK{=}16384$ unchanged.
This preserves the same AIC throughput but reduces the $A$-side footprint $MK$, thereby improving compute efficiency.
As a result, $(128,256,64)$ achieves the same AIC throughput as $(128,128,128)$ with lower input traffic.
Specifically, the fp16 input traffic per tile equals the combined footprints of the $A$- and $B$-tiles, i.e., $2(MK+NK)$ bytes.
Thus, $(128,256,64)$ transfers $2(128\!\cdot\!64+256\!\cdot\!64)=48$\,KB, whereas $(128,128,128)$ transfers $2(128\!\cdot\!128+128\!\cdot\!128)=64$\,KB.

\Paragraph{Final Tile Choice.}
The above trade cannot be extended indefinitely, since the fp32 output must satisfy $MN \le 32768$ under double-buffered L0C, which bounds $N \le 256$ when $M=128$.
Combining AIC utilization, input traffic, buffer residency, and write-back efficiency, we adopt
\[
(M,N,K) = (128,256,64)
\]
as the default tile configuration.
This choice maximizes effective compute density while respecting all on-chip constraints, enabling sustained intra-core tile reuse.

\section{Implementation}

\begin{figure}[t]
\centering

\begin{minipage}[t]{0.48\linewidth}
\centering
\includegraphics[width=\linewidth]{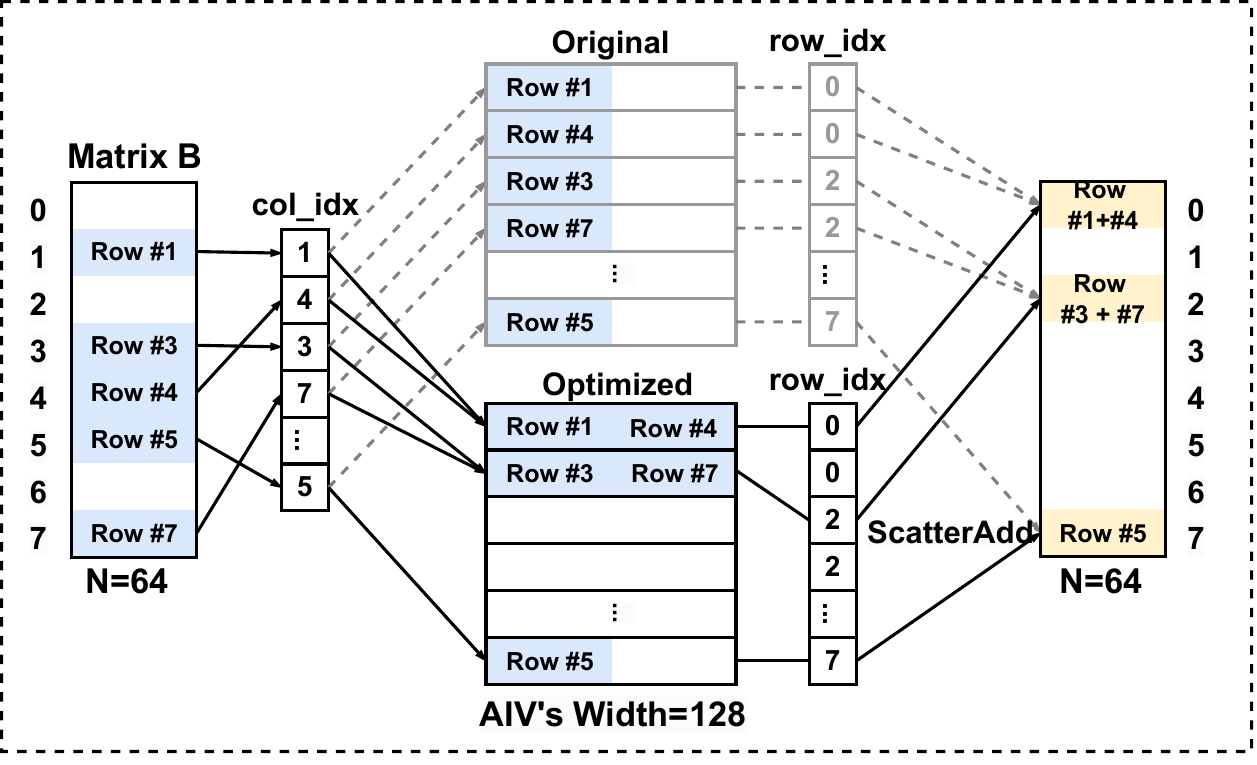}
\vspace{-0.1in}
\captionof{figure}{Vector tiles merging.}
\label{AIV_Kernel}
\end{minipage}
\hfill
\begin{minipage}[t]{0.48\linewidth}
\centering
\includegraphics[width=\linewidth]{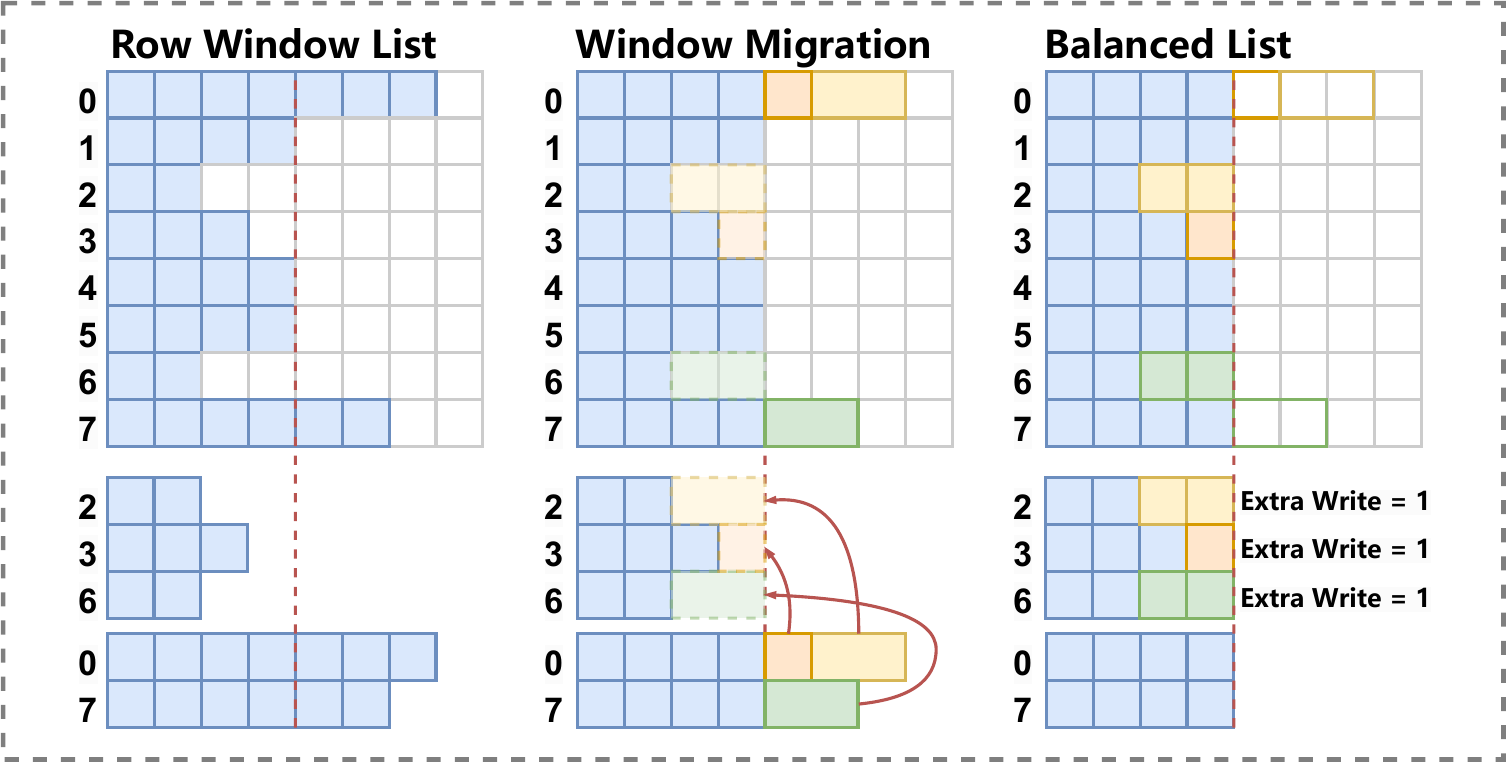}
\vspace{-0.1in}
\captionof{figure}{Row window list reordering.}
\label{AIC_Kernel}
\end{minipage}

\vspace{-0.1in}
\end{figure}

\Paragraph{Vector Tiles Merging (AIV-side).}
Figure~\ref{AIV_Kernel} illustrates vector tiles merging in the AIV runtime.
In sparse SpMM, AIV repeatedly gathers rows of $B$ by \texttt{col\_idx} and accumulates them to output rows by \texttt{row\_idx}.
When $N$ (the column width of dense matrix $B$) is not divisible by AIV's 128-wide vector width, each gather-accumulate activates only a fraction of SIMD lanes, wasting both compute and bandwidth.
The optimization groups entries sharing the same \texttt{row\_idx} and packs multiple gathered $B$ rows into one widened vector tile (Figure~\ref{AIV_Kernel}, optimized), concatenating them to reach a 128-aligned shape, so a single AIV step processes a denser tile and then performs one \textit{ScatterAdd} to the target output row, improving lane utilization and reducing fragmented memory accesses.



\Paragraph{Row Window List Migration (AIC-side).}
Figure~\ref{AIC_Kernel} illustrates row window list migration in the AIC runtime. After tile preparation, computation is organized as a row window list, where each row window corresponds to one input tile and is assigned to an AIC. A naive row-wise assignment can be highly imbalanced because different lists contain different numbers of row windows, leading to long-tail stragglers. The migration step reshuffles row windows across the list to form a balanced list (Figure~\ref{AIC_Kernel}, middle), interleaving heavy and light windows without splitting windows or changing tile shapes. This incurs one extra write to reconcile the migrated windows with their original output locations, but it substantially reduces stragglers and improves throughput.

\Paragraph{Double Buffer Pipelining.}
We implement double buffer pipelining in both AIC and AIV to overlap computation with data movement.
Each engine alternates between two buffer sets, allowing one tile to execute while the next tile is being prefetched into on-chip buffers or written back.
This design hides data transfer latency behind computation and enables continuous tile execution, at the cost that each tile may occupy at most half of the local buffer capacity.

\section{Evaluation}
\label{sec7}

%

\subsection{Experimental Setup}

\Paragraph{Environments.} The experiments are conducted on two servers: an NPU server equipped with an Ascend 910B (24 AI Cores and 64 GB HBM) enabled with CANN v8.0.1 runtime, and a GPU server equipped with an NVIDIA A100 (Ampere, 108 SMs each consisting of 64 CUDA Cores and 4 Tensor Cores, 80 GB HBM) enabled with CUDA v12.3. The server runs Ubuntu 18.04 with Linux kernel 4.13.0.

\Paragraph{Datasets.} We use four real-world GNN benchmarks (cora \cite{mccallum2000cora}, ogbn-arxiv \cite{hu2020ogbn-arxiv}, reddit \cite{hamilton2017reddit}, and amazon-product \cite{mcauley2015amazon}) and sixteen sparse matrices from the SuiteSparse \cite{davis2011suitesparse}. Table~\ref{tab:datasets} lists the datasets. ``Abbr.'' denotes dataset abbreviations, ``Row\&Col'' gives the row and column dimensions of the input sparse matrix ($M \times K$), ``NNZ'' is the number of nonzeros, and ``AvgL'' indicates the average row length. 
Table~\ref{tab:datasets} summarizes the datasets and their key structural characteristics. “Abbr.” denotes dataset abbreviations, “Row\&Col” gives the matrix dimensions, “NNZ” is the number of nonzeros, “AvgL” is the average row length, \zyf{“Dens.” reports the matrix density scaled by $10^{-3}$, “Skew” measures the fraction of NNZ contained in the top 10\% rows ranked by row NNZ, and “Empty” reports the fraction of empty 16$\times$16 tiles before tile reordering.}



\begin{table}[t]\small
\centering
\caption{\zyf{Datasets}}
\label{tab:datasets}
\setlength{\tabcolsep}{3pt}
\scriptsize
\begin{tabular}{c c c c c @{\hspace{2pt}} c @{\hspace{2pt}} c @{\hspace{2pt}} c}
\toprule
Dataset & Abbr. & Row\&Col & NNZ & AvgL & \zyf{Dens. ($10^{-3}$)} & \zyf{Skew} & \zyf{Empty Tiles} \\
\hline
cora            & CR  & 2,708     & 10,556      & 3.9    & 1.44  & 0.32 & 0.05 \\
wiki-RfA        & WR  & 11,380    & 362,053     & 31.8   & 2.80  & 0.39 & 0.01 \\
dawson5         & DA  & 51,537    & 1,010,777   & 19.6   & 0.38  & 0.14 & 0.98 \\
olafu           & OL  & 16,146    & 1,015,156   & 62.9   & 3.89  & 0.12 & 0.99 \\
ogbn-arxiv      & OA  & 169,343   & 2,315,598   & 13.6   & 0.08  & 0.50 & 0.86 \\
pattern1        & PA  & 19,242    & 9,323,432   & 484.5  & 25.18 & 0.16 & 0.73 \\
mip1            & MP  & 66,463    & 10,352,819  & 155.8  & 2.34  & 0.17 & 0.99 \\
mycielskian15   & M15 & 24,575    & 11,111,110  & 452.1  & 18.40 & 0.42 & 0.89 \\
nd12k           & ND  & 36,000    & 14,220,946  & 395.0  & 10.97 & 0.12 & 0.95 \\
human\_gene1    & HG  & 22,283    & 24,669,643  & 1107.1 & 49.68 & 0.24 & 0.49 \\
F1              & F1  & 343,791   & 26,837,113  & 78.1   & 0.23  & 0.44 & 0.98 \\
ML\_Laplace     & ML  & 377,002   & 27,689,972  & 73.4   & 0.19  & 0.10 & 0.99 \\
Fault\_639      & FA  & 638,802   & 28,614,564  & 44.8   & 0.07  & 0.12 & 0.99 \\
mouse\_gene     & MG  & 45,101    & 28,967,291  & 642.3  & 14.24 & 0.41 & 0.50 \\
audikw\_1       & AU  & 943,695   & 77,651,847  & 82.3   & 0.09  & 0.24 & 0.99 \\
mycielskian17   & M17 & 98,304    & 100,245,742 & 1019.8 & 10.37 & 0.46 & 0.94 \\
reddit          & RD  & 232,965   & 114,615,892 & 492.0  & 2.11  & 0.46 & 0.01 \\
amazon-product  & AP  & 2,449,029 & 123,718,152 & 50.5   & 0.02  & 0.45 & 0.96 \\
mycielskian18   & M18 & 196,607   & 300,933,832 & 1530.6 & 7.79  & 0.48 & 0.95 \\
mycielskian19   & M19 & 393,215   & 903,194,710 & 2296.9 & 5.84  & 0.50 & 0.96 \\
\bottomrule
\end{tabular}
\end{table}

\Paragraph{Competitor baselines.} In performance evaluation, we compare \system against representative SpMM baselines on both Ascend NPUs and GPUs. \aix{On NPUs, we include MindSporeGL \cite{zhuPerformanceEvaluationMindSpore2023,Ascend_HPCA_2021}, which mainly executes sparse kernels on AIV, and an AIC-based SpMM design reimplemented from prior Ascend work \cite{moustafa2023accelerating}, which primarily relies on AIC execution.}    
On GPUs, we use cuSPARSE \cite{nvidia2023cusparse}, NVIDIA’s vendor-optimized sparse computation library with deeply tuned SpMM and SpMV kernels, DTC-SpMM \cite{fanDTCSpMMBridgingGap2024}, a SOTA SpMM framework that exploits Tensor Cores, and HC-SpMM \cite{liHCSpMMAcceleratingSparse2024}, a recent GPU SpMM framework that jointly leverages CUDA Cores and Tensor Cores to improve utilization under sparsity. By default, we use $N=256$ as the column width of the dense matrix $B$, unless stated otherwise.




\begin{figure}[t]
  \centering
  \includegraphics[width=0.7\linewidth]{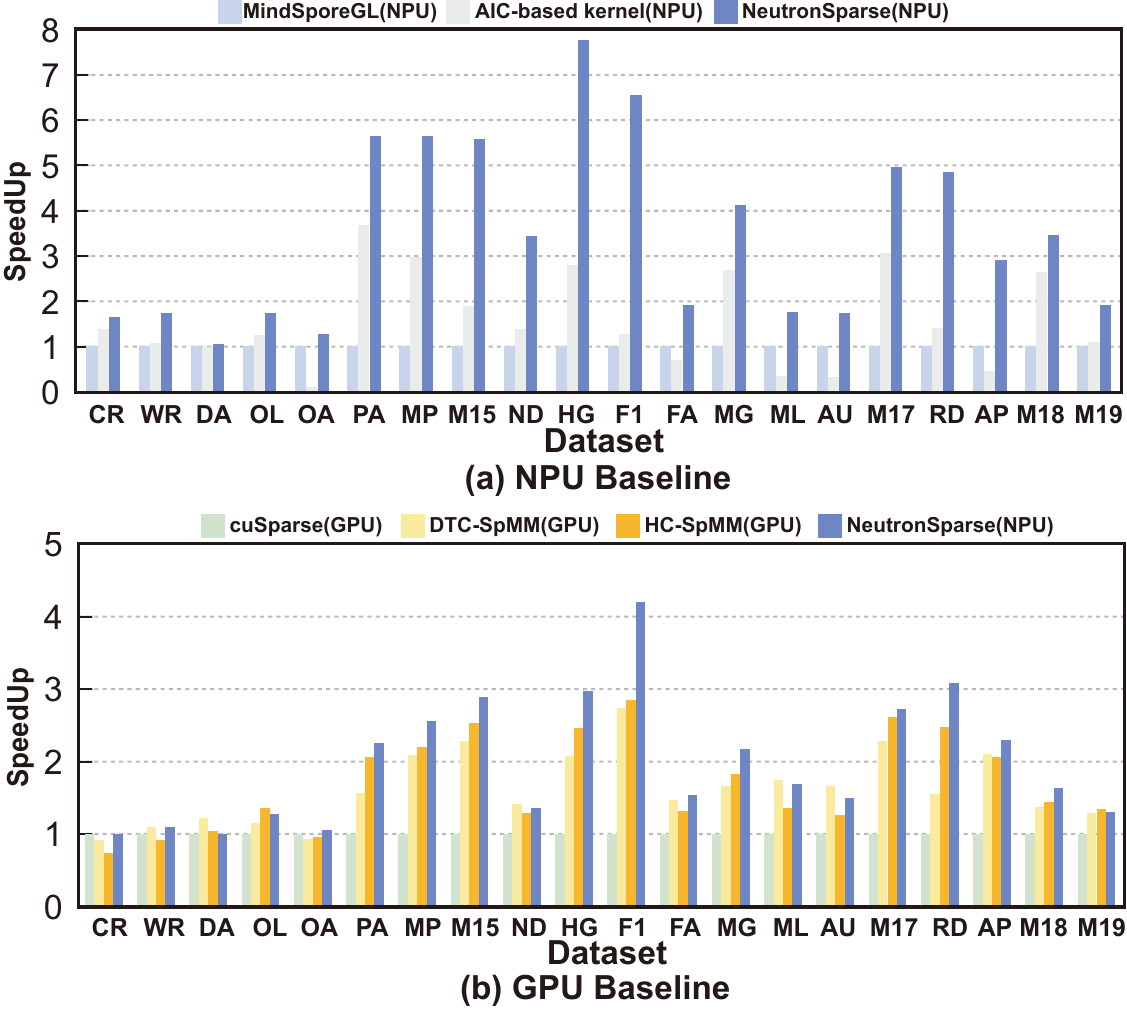}
  \vspace{-0.1in}
  \caption{{SpMM performance on Ascend 910B and Nvidia A100}}
  \label{Experiment of NPU and GPU SpeedUp}
  \vspace{-0.1in}
\end{figure}
\subsection{Overall Comparison}

We conduct a comprehensive performance comparison of \system, MindSporeGL, cuSPARSE, DTC-SpMM, and HC-SpMM. Figure~\ref{Experiment of NPU and GPU SpeedUp} shows the results: on Ascend NPUs, speedups are normalized to MindSporeGL, while on GPUs, speedups are normalized to cuSPARSE.

\Paragraph{Comparison with the NPU baseline.}  
Compared to both MindSporeGL and the AIC-based baseline, \system consistently achieves superior performance, with an average speedup of $3.48\times$ over MindSporeGL and $3.25\times$ over the AIC-based design.
\aix{The two baselines expose complementary limitations: MindSporeGL relies solely on AIV, leaving AIC underutilized, while the AIC-based design centers on dense-tile execution, incurring substantial redundant computation on zero entries and executing AIC and AIV largely in a serial manner. In contrast, \system jointly exploits AIC and AIV through fine-grained partitioning and coordinated execution, enabling both high compute efficiency and effective overlap.}
The gains are particularly pronounced on higher-density matrices (e.g., \texttt{human\_gene1}), where leveraging AIC becomes critical and \system achieves up to $7.78\times$ speedup.



\Paragraph{Comparison with the GPU baseline.}
Compared to cuSPARSE, DTC-SpMM, and HC-SpMM, \system exhibits superior performance across most datasets, achieving an average speedup of $1.98\times$ (up to $4.19\times$), $1.27\times$ (up to $1.99\times$), and $1.18\times$ (up to $1.47\times$), respectively.
cuSPARSE and DTC-SpMM mainly rely on a single dominant execution engine (CUDA cores or Tensor Cores), which performs well only on the regions that match its compute pattern.
Although HC-SpMM combines CUDA cores and Tensor Cores to cover different workload types, its kernels are still executed in a largely staged manner and cannot concurrently run heterogeneous engines to maximize sustained throughput.
In contrast, \system can overlap sparse processing and dense tile computation on the NPU’s heterogeneous engines via online coordination, thus handling mixed sparsity patterns more effectively. DTC-SpMM and HC-SpMM achieve better performance on cora, wiki-RfA, and dawson5. Because these datasets are small in scale, the benefits of \system’s coordinating execution are insufficient to outweigh the additional scheduling overhead.

\subsection{Analysis of AIV-AIC Coordination}

\begin{figure}[t]
  \centering
   \vspace{-0.1in}
  \includegraphics[width=0.7\linewidth]{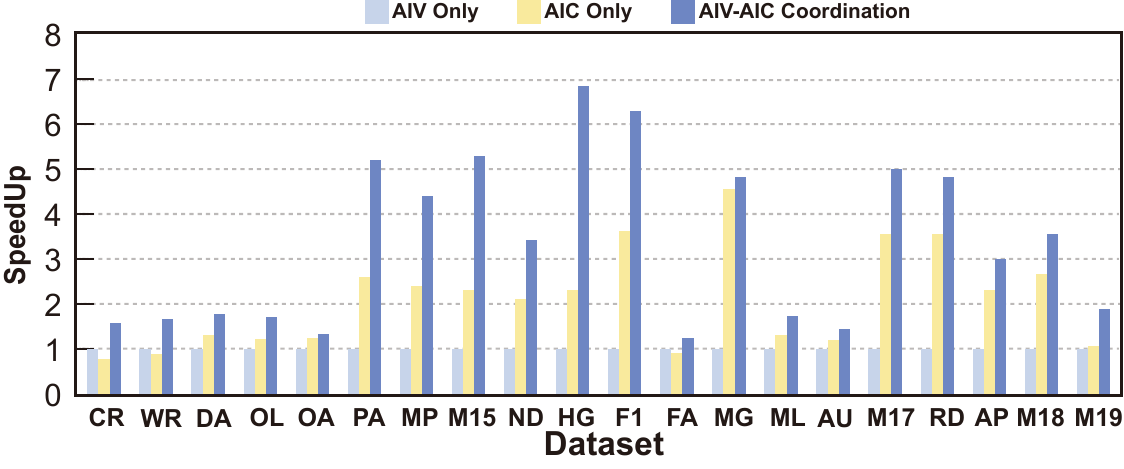}
   \vspace{-0.1in}
  \caption{Performance gain of AIV-AIC Coordination.}
  \label{ablation_experiment_unit}
\end{figure}

\Paragraph{Performance Gain.} To evaluate the performance benefits of our proposed AIV-AIC coordinating optimization for SpMM, we use 15 datasets of varying scales and sparsity. We compare three kernels: (1) AIV-only: a baseline kernel that exclusively utilizes AIV; (2) AIC-only: a baseline kernel using only AIC; and (3) AIV-AIC Coordination: the hybrid heterogeneous kernel applied in \system.
Figure~\ref{ablation_experiment_unit} shows the normalized speedup with the AIV-only kernel as the baseline. AIV-AIC Coordination achieves the best performance on all test matrices, with average speedups of $3.71\times$ and $1.98\times$ over the AIV-only and AIC-only kernels, respectively. 
The largest speedups are observed on \texttt{human\_gene1} and \texttt{F1}, whose power-law sparsity induces clear separation between sparse and dense regions.
This heterogeneity amplifies the benefit of AIV–AIC coordination, allowing \system to map sparse work to AIV and dense tiles to AIC more effectively.

\begin{figure}[t]
  \centering
  \includegraphics[width=0.9\linewidth]{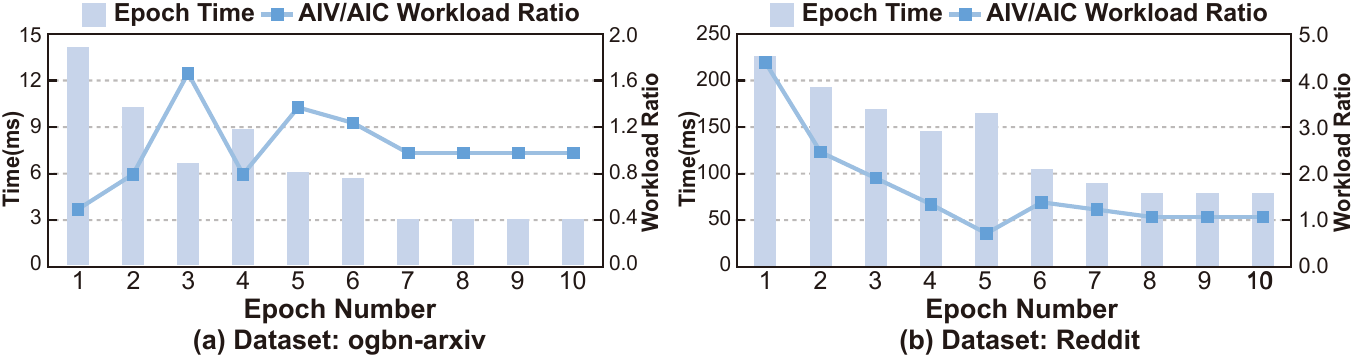}
  \vspace{-0.1in}
  \caption{{Dynamic balance AIV-AIC workload.}}
  \label{DynamicWorkloadBalance}
  \vspace{-0.1in}
\end{figure}

\Paragraph{Online Workload Migration.}
Figure~\ref{DynamicWorkloadBalance} shows the online workload migration between AIV and AIC on two datasets.
\system monitors the real-time progress of the two engines in the coordinated pipeline and migrates tiles from the slower engine to the faster one to eliminate stragglers.
Starting from the initial static partition, the AIV/AIC workload ratio quickly converges to a stable range within the first 10 epochs, and the per-epoch time drops accordingly, yielding an average speedup of $3.83\times$ over the initial execution time after convergence.
\zyf{The workload ratio obtained via online migration is not strictly optimal, but is already close to the optimal time-balanced solution. Specifically, migration stops once the execution time difference between AIV and AIC falls below a small tolerance (e.g., $5\%$), which bounds the final ratio within a near-optimal region. Starting from the optimal ratio could further improve performance, but the gain would be limited. Since the final ratio is already tightly bounded and the migration overhead is small compared to the overall runtime (e.g., 5\% among hundreds of epochs), the additional improvement is marginal (up to $10\%$).}
By adapting the partition online, \system sustains time-balanced progress between AIV and AIC and achieves stable throughput improvements.

\begin{figure}[t]
  \centering
  \includegraphics[width=0.9\linewidth]{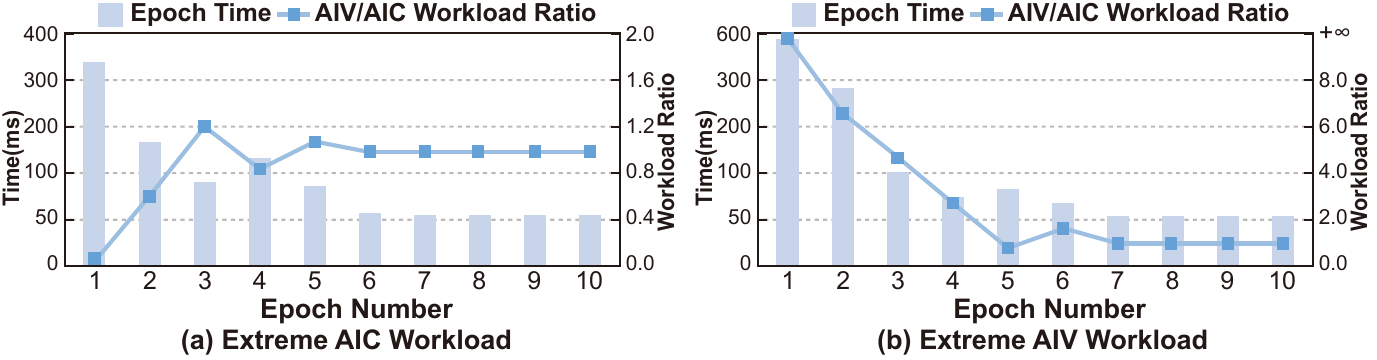}
  \vspace{-0.1in}
  \caption{{\zyf{Convergence of online workload migration under extreme initial-skew cases.}}}
  \label{Extreme_Workload}
  \vspace{-0.1in}
\end{figure}

We further construct extreme-skew cases from \texttt{reddit} by artificially adjusting the number of nonzeros in each row, so that the initial workload is assigned almost entirely to either AIC or AIV. Even under such extreme imbalance, online migration still converges quickly. This process can be viewed as a bisection-style iterative balancing procedure on the residual imbalance: each round repartitions the remaining workload according to the observed execution-time skew, so the imbalance shrinks geometrically and the number of adjustment rounds grows only logarithmically with the initial skew. As shown in Fig.~\ref{Extreme_Workload}, the adjustment converges within 7 rounds, only slightly slower than on real workloads despite the highly imbalanced initialization. Notably, we observe that the final stable ratio in case (b) is AIC-dominated because AIC provides significantly higher effective compute throughput than AIV, even when the raw sparsity pattern appears more suitable for vector-style execution. As a result, after balancing execution time, a larger fraction of workload is assigned to AIC to fully utilize its higher processing capacity.

\begin{figure}[t]
  \centering
  \includegraphics[width=0.9\linewidth]{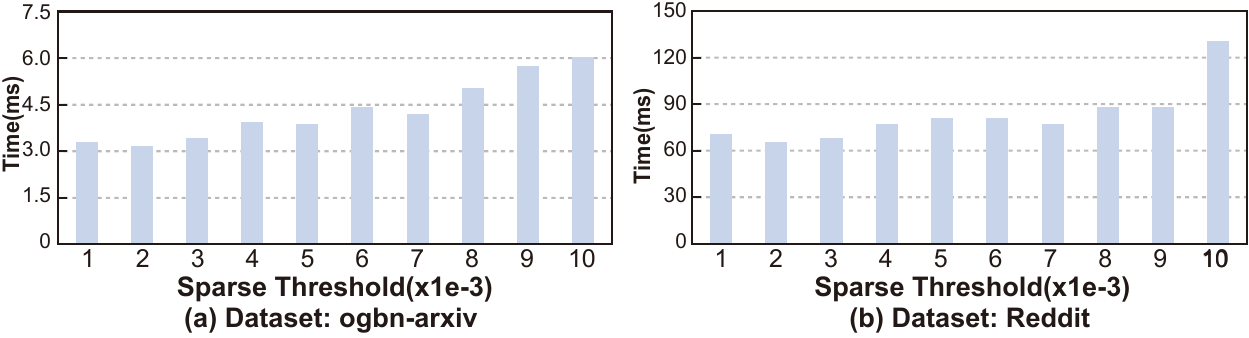}
  \vspace{-0.1in}
  \caption{{\wqg{Sensitivity of SpMM performance to the initial sparsity threshold.}}}
  \label{ThresholdAnalysis}
  \vspace{-0.1in}
\end{figure}

\Paragraph{\wqg{Sensitivity to the initial sparsity threshold}}
\wqg{We further study the sensitivity of heterogeneous workload partitioning to the initial sparsity threshold. In our implementation, this threshold determines the initial AIV/AIC split: rows with density above the threshold are assigned to AIC, while the others are assigned to AIV. Since the densities of the target sparse matrices are mostly on the order of $10^{-3}$, we sweep the threshold from $1\times 10^{-3}$ to $10\times 10^{-3}$ and evaluate the resulting performance on \texttt{ogbn-arxiv} and \texttt{reddit}. As shown in Fig.~\ref{ThresholdAnalysis}, the default threshold $2\times 10^{-3}$ already lies near the optimal-performing region. In particular, thresholds from $1\times 10^{-3}$ to $3\times 10^{-3}$ lead to only 6.4\% performance variation on average, while larger deviations cause more noticeable degradation due to a worse initial workload balance. This result suggests that the lightweight cost model is sufficiently accurate to provide a good starting point, while the remaining mismatch can be corrected by online migration.}

\subsection{Analysis of Tile Orchestrating}

\begin{figure}[t]
  \centering
  \includegraphics[width=0.7\linewidth]{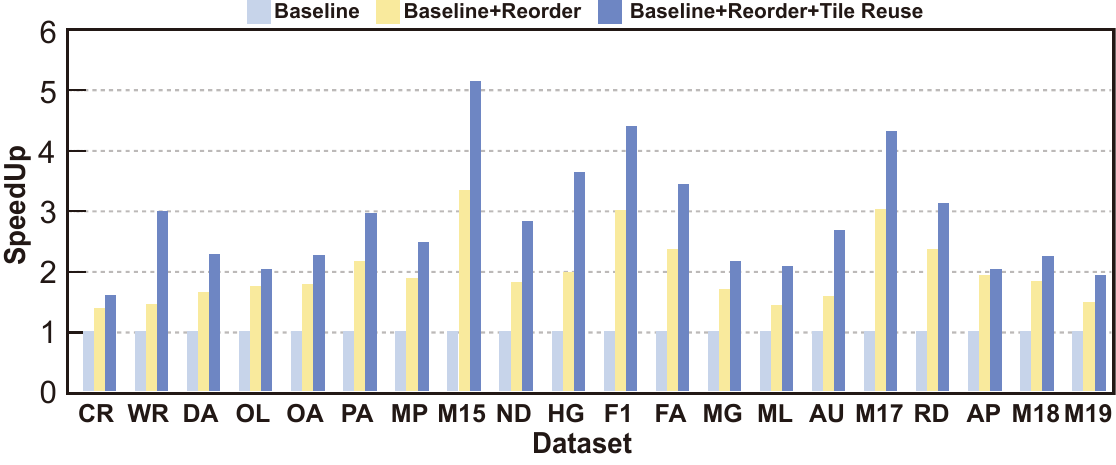}
   \vspace{-0.1in}
  \caption{{Performance gain of locality-aware tile scheduling.}}
  \label{experiment_TileReuse}
   \vspace{-0.1in}
\end{figure}

\Paragraph{Performance Gain.}
We quantify the benefit of the two techniques in Section~6 via an ablation study on 15 datasets (Figure~\ref{experiment_TileReuse}). 
The baseline is a plain tile-based SpMM using a fixed tile shape $(M,N,K)=(128,128,64)$; we then progressively enable global-local reordering (Baseline+Reorder) and further add hierarchical tile reusing ({Baseline+Reorder+Tile Reuse}), reporting speedups normalized to the baseline.
Overall, {Baseline+Reorder} achieves an average speedup of $2.17\times$ by clustering correlated rows/columns into communities and forming denser row-window tiles, which reduces redundant work on sparsely populated tiles. 
On top of that, {Tile Reuse} provides an additional $1.42\times$ speedup over {Baseline+Reorder} by reducing repeated gathers of dense $B$ rows and increasing per-core tile residency so that loaded data is reused across more MAC operations before being evicted.
The larger gains appear on datasets such as {F1} and {M15}, which exhibit strong power-law sparsity with pronounced structural skew; this makes our global-local reordering particularly effective at clustering highly correlated rows/columns and producing much denser row-window tiles.

\begin{figure}[t]
  \centering
  \includegraphics[width=0.7\linewidth]{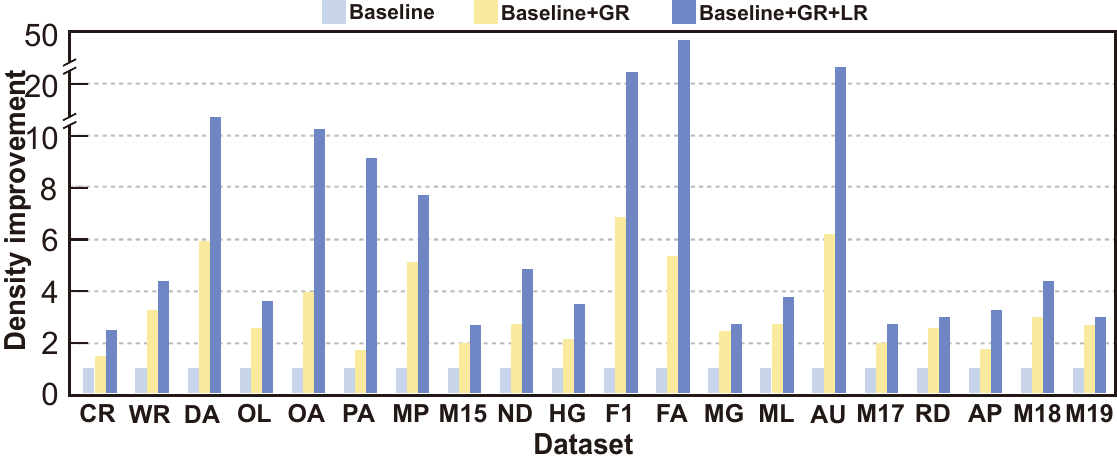}
   \vspace{-0.1in}
  \caption{Density improvement of global-local reordering. "GR": global reordering, "LR": local reordering.}
  \label{ablation_experiment_reordering}
   \vspace{-0.1in}
\end{figure}

\Paragraph{Impact on Tile Density.} We analyze the effectiveness of Global-local Reordering on AIC workloads, including Global Reordering (GR) and Local Reordering (LR).
We use {density improvement} as the metric, defined as the multiplicative increase in the nonzero ratio after reordering, i.e.,
$\textit{DensityImprovement} \!=\! \rho_{\text{after}} / \rho_{\text{before}}$, where $\rho=\text{NNZ}/(M\!\cdot\!K)$ is the tile density.
We start from an unoptimized AIC workload and progressively enable GR and LR.
Figure~\ref{ablation_experiment_reordering} reports the normalized density improvement.
Compared with the baseline, baseline+GR increases tile density by clustering correlated rows/columns via community detection, achieving an average $3.44\times$ improvement.
Building on baseline+GR, baseline+GR+LR further boosts density by {reordering rows within each cluster} to better align nonzeros across columns, and then {compacting away empty columns} during tile construction, achieving an average $10.06\times$ improvement.
GR and LR show larger improvements on F1, Fault\_639, and audikw\_1 because these matrices have shorter average row lengths and thus contain more zero-only regions that can be eliminated by clustering and compaction.
In contrast, pattern1 benefits less since its higher baseline density leaves limited room for further compression.

\begin{figure}[t]
  \centering
  \includegraphics[width=0.7\linewidth]{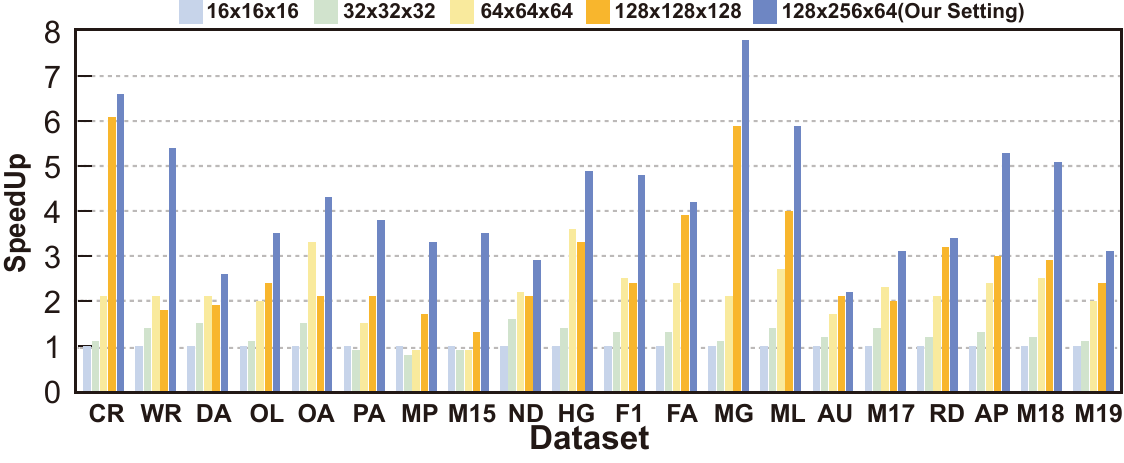}
   \vspace{-0.1in}
  \caption{{\wqg{Performance with varying tile sizes.}}}
   \label{tilesize}
   \vspace{-0.1in}
\end{figure}

\Paragraph{\wqg{Performance with Varying Tile Sizes.} }
\wqg{Tile shape has a clear impact on overall performance because it directly affects compute efficiency, data reuse, and buffer utilization. To evaluate this effect, we compare our default tile shape $(128,256,64)$ with four alternatives: $(16,16,16)$, $(32,32,32)$, $(64,64,64)$, and $(128,128,128)$. The results show that $(128,256,64)$ consistently achieves the best performance, outperforming these alternatives by $3.39\times$, $2.84\times$, $1.64\times$, and $1.31\times$ on average, respectively. Smaller tiles suffer from lower compute intensity and weaker data reuse, while larger symmetric tiles are less effective in balancing on-chip buffer usage and write-back efficiency. These results validate that our default tile shape provides a better trade-off among compute-to-memory ratio, buffer constraints, and output efficiency.}


\subsection{Scalability Analysis}
\begin{figure}[t]
  \centering
  \includegraphics[width=0.7\linewidth]{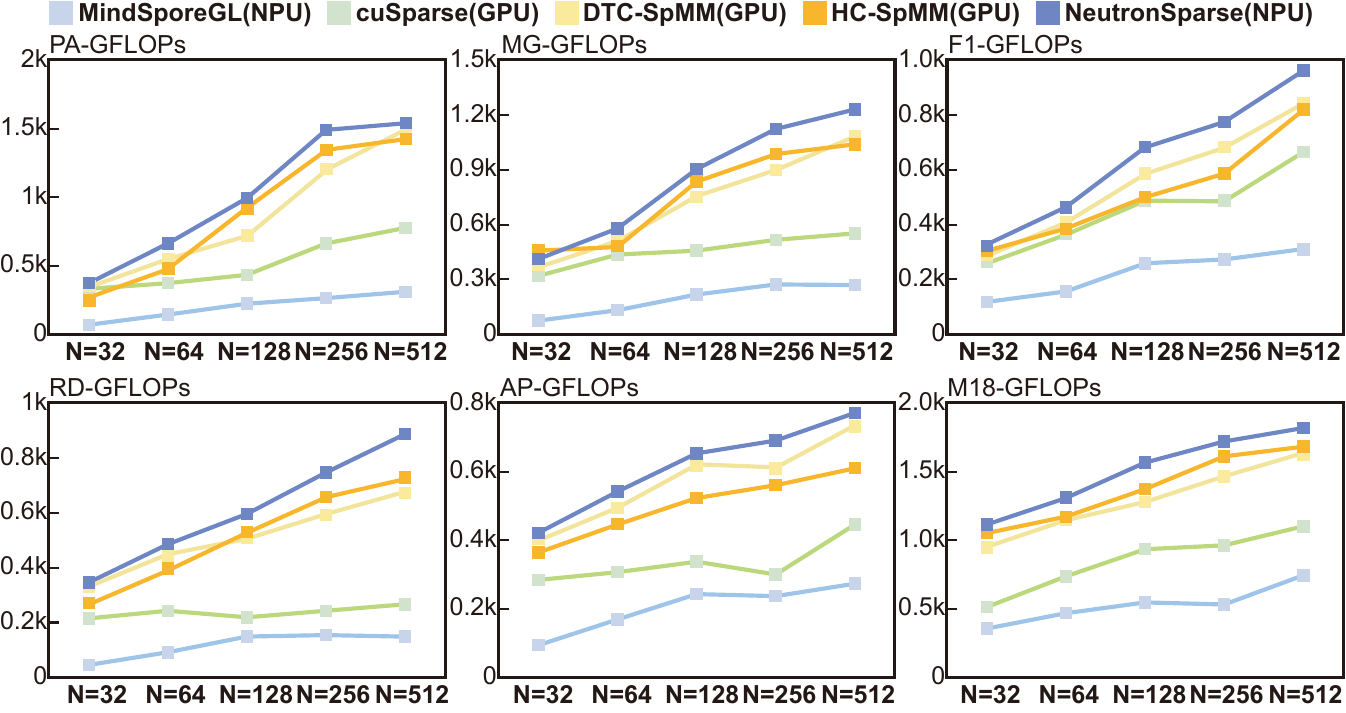}
  \vspace{-0.1in}
  \caption{Scaling performance when varying matrix sizes.}
  \label{Experiment of GFLOPS}
  \vspace{-0.1in}
\end{figure}

In this experiment, we compare \system with all baselines with different column widths $N$ (set to $\{32, 64, 128, 256, 512\}$) of matrix $B$ over six datasets (PA, MG, F1, RD, AP, M18). Figure~\ref{Experiment of GFLOPS} shows the results. Across different N sizes, \system consistently outperforms the baselines. Specifically, as 
N increases from 32 to 512, {\system achieves $1.35\times$–$4.04\times$ GFLOPs improvement, MindSporeGL $1.18\times$–$2.09\times$, cuSPARSE $1.26\times$–$2.38\times$, DTC-SpMM $1.31\times$–$3.68\times$, and HC-SpMM $1.32\times$-$3.88\times$}.
We observe that \system, DTC-SpMM, and HC-SpMM exhibit higher GFLOPs improvement than cuSPARSE and MindSporeGL.
That is because the larger N raises the compute-to-memory ratio, thereby amplifying the computation bottlenecks of cuSPARSE and MindSporeGL, whose advantages mainly lie in memory access. In contrast, \system, DTC-SpMM, and HC-SpMM employ matmul-based units, where the higher compute-to-memory ratio helps alleviate the memory access bottleneck of density computing units.

\subsection{Preprocessing Overhead Analysis}

\zyf{We next analyze the preprocessing overhead introduced by \system. Since this overhead is mainly determined by dataset scale, we use representative small-, medium-, and large-scale datasets to present the trend clearly.}

\begin{table}[t]\small
  \centering
  \caption{Runtime breakdown of 200-epoch GNN training (seconds and percentage of total time).}
  \vspace{-0.1in}
  \label{tab:overhead_breakdown}
  \scriptsize
  \begin{tabular}{l c c c c}
    \toprule
    \multirow{2}{*}{\raisebox{-0.3ex}{\textbf{Dataset}}} &
    \multicolumn{3}{c}{\textbf{Runtime Breakdown (Ratio)}} &
    \multirow{2}{*}{\raisebox{-0.3ex}{\textbf{Total (s)}}} \\
    \cmidrule(lr){2-4}
    & \textbf{Partition} & \textbf{Reorder} & \textbf{GCN Exec.} & \\
    \midrule
    \texttt{cora}           & 0.01 (2.38\%)   & 0.018 (4.29\%)  & 0.392 (93.33\%)   & 0.42   \\
    \texttt{ogbn-arxiv}     & 0.016 (2.35\%)  & 0.015 (2.21\%)  & 0.649 (95.44\%)   & 0.68   \\
    \texttt{reddit}         & 9.34 (3.48\%)   & 6.32 (2.35\%)   & 252.84 (94.17\%)  & 268.5  \\
    \texttt{amazon-product} & 10.37 (3.12\%)  & 9.62 (2.90\%)   & 311.81 (93.98\%)  & 331.8  \\
    \texttt{mycielskian18}  & 15.88 (3.87\%)  & 10.34 (2.52\%)  & 384.08 (93.61\%)  & 410.3  \\
    \bottomrule
  \end{tabular}
\end{table}

\Paragraph{Amortized Overhead Analysis.} 
We evaluate the amortized overhead of the two preprocessing steps introduced by \system, heterogeneous workload partitioning and global-local reordering, under long-running training workloads. Specifically, we run 200 epochs of GCN training and report the runtime breakdown in Table~\ref{tab:overhead_breakdown}. Both preprocessing steps are executed once before training and their costs are amortized across all epochs. On average, heterogeneous workload partitioning accounts for approximately 3.04\% of the total runtime, while global-local reordering contributes about 2.85\%. In contrast, GNN training is dominated by SpMM execution, which consistently accounts for over 93\% of the end-to-end runtime across all datasets. These results indicate that the overhead of our preprocessing is negligible while enabling substantial performance benefits during iterative SpMM execution.

\Paragraph{Overhead Comparison.}
We compare the preprocessing overhead of \system and DTC-SpMM on \texttt{cora}, \texttt{ogbn-arxiv}, and \texttt{reddit}.
Table~\ref{tab:overhead_comparison} presents the results.
DTC-SpMM relies on a global row--column reordering over the entire sparse matrix, which scans and permutes all rows and columns and thus incurs substantial preprocessing cost.
In contrast, the lower overhead of \system stems from two design choices.
First, heterogeneous workload partitioning identifies and assigns extremely sparse rows and columns (with very few nonzeros) to the AIV stream in a single pass, completely bypassing reordering for this portion of the matrix.
Second, for the remaining AIC workload, reordering is confined to two lightweight stages: a coarse cluster-level grouping followed by row-only reordering within each cluster, rather than a full global row--column permutation.
As a result, \system consistently incurs significantly lower preprocessing overhead than DTC-SpMM, reducing preprocessing cost by $4.25\times$ up to $156.92\times$, with the advantage becoming more pronounced on larger datasets where global reordering grows increasingly expensive.

\begin{table}[t]\small
  \centering
  \caption{Preprocessing overhead comparison.}
  \vspace{-0.1in}
  \label{tab:overhead_comparison}
  \scriptsize
  \begin{tabular}{@{} l c c c @{}}
    \toprule
    \multirow{2}{*}{\raisebox{-0.3ex}{\textbf{Dataset}}} &
    \multirow{2}{*}{\raisebox{-0.3ex}{\textbf{Rows\&Col}}} &
    \multicolumn{2}{c}{\textbf{Overhead}} \\
    \cmidrule(lr){3-4}
    & & \textbf{\system} & \textbf{DTC-SpMM} \\
    \midrule
    \texttt{cora}            & 2,708        & 0.4s   & 1.7s    \\
    \texttt{ogbn-arxiv}      & 169,343      & 0.7s   & 12.3min \\
    \texttt{amazon-product}  & 1,569,960    & 6.5min & 17 h    \\
    \bottomrule
  \end{tabular}
\end{table}


\subsection{Power Consumption Analysis}

\begin{figure}[t]
  \centering
  \includegraphics[width=0.7\linewidth]{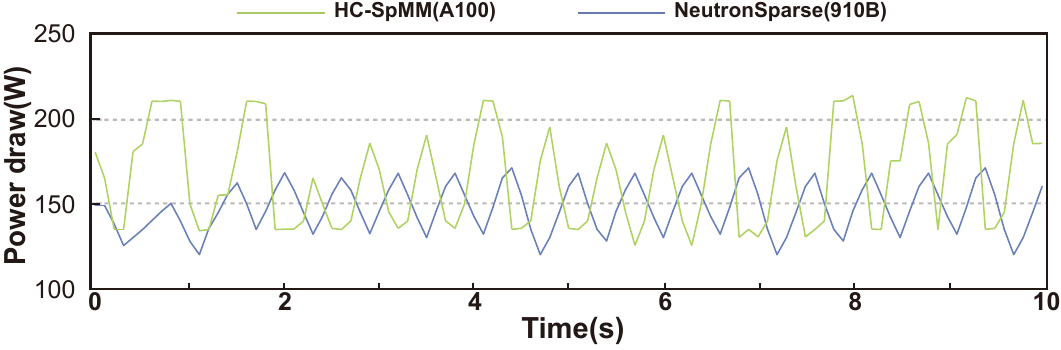}
  \vspace{-0.1in}
  \caption{{Power draw (in watts) sampled every 100ms over a 10-second training window (Dataset: reddit).}}
  \label{fig:compare-power}
  \vspace{-0.1in}
\end{figure}

To evaluate energy efficiency, we record the power draw (in watts) every 100 milliseconds over a 10-second time window during SpMM. On Ascend, power draw is collected using \texttt{npu-smi}, while on GPUs, \texttt{nvidia-smi} is used. 
Figure \ref{fig:compare-power} shows the results. \system, running on the Ascend 910B, achieves the lower power consumption, with an average power draw of 147.7W and a peak of 171.2W. In contrast, HC-SpMM incurs higher consumption, with an average power draw of 166.31W and a peak of 213.7W.
The consistent gap across platforms indicates that \system achieves competitive performance and resource efficiency with lower power consumption.


\begin{figure}[t]
  \centering
 \includegraphics[width=0.7\linewidth]{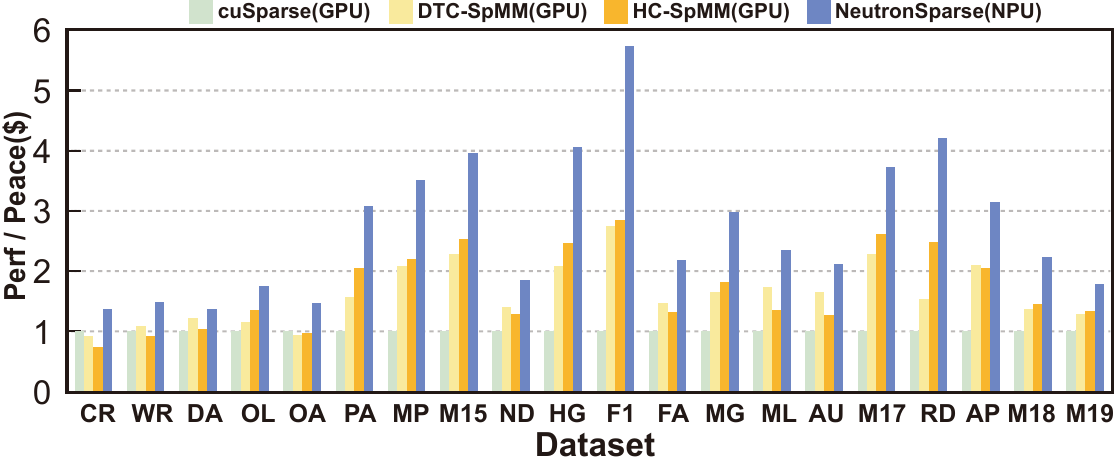}
  \vspace{-0.15in}
  \caption{{Cost-normalized SpMM throughput using hourly cloud rental prices.}}
  \label{fig:cost_efficiency}
  \vspace{-0.1in}
\end{figure}

\subsection{Cost Efficiency Analysis}
We evaluate cost efficiency as throughput normalized by hourly rental price. \aix{To reduce potential bias from pricing variability, we collect rental prices from multiple mainstream cloud platforms in the Asia-Pacific region, including Huawei Cloud, AutoDL, Parallel Technology, Microsoft Azure, AWS EC2, and Google Cloud, and use the averaged price for each device type.} Figure~\ref{fig:cost_efficiency} reports the normalized results.
After price normalization, \system consistently outperforms GPU-based baselines in performance per dollar. On average, \system achieves $2.47\times$, $1.49\times$, and $1.43\times$ higher cost efficiency than cuSPARSE, DTC-SpMM, and HC-SpMM, respectively. These results indicate that the advantage mainly comes from improved execution efficiency rather than pricing differences, suggesting that NPU-aware optimization can translate hardware capability into tangible cost benefits.

\begin{figure}[t]
  \centering
  \includegraphics[width=0.7\linewidth]{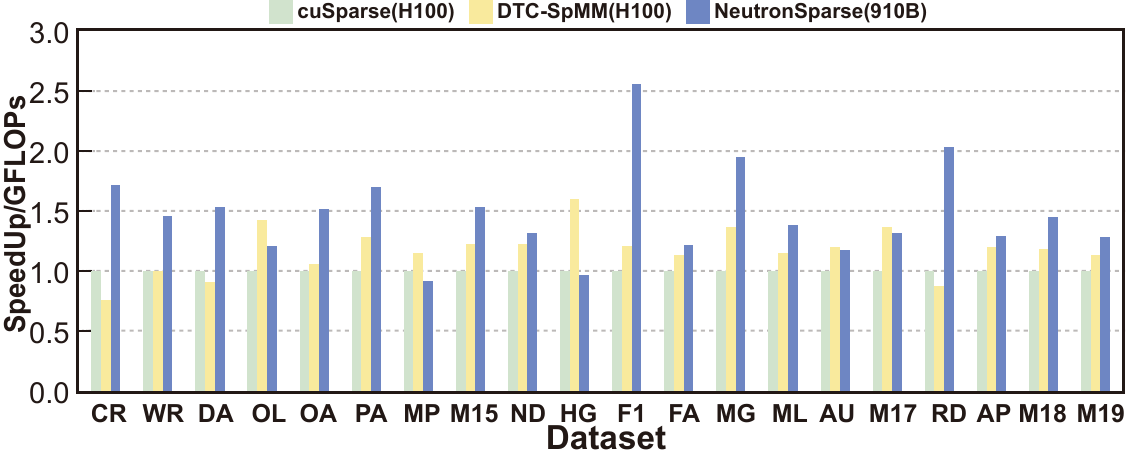}
  \vspace{-0.1in}
  \caption{{\wqg{TFLOPS-normalized SpMM performance comparison on Ascend 910B and NVIDIA H100.}}}
  \label{fig:compare-H100}
  \vspace{-0.1in}
\end{figure}

\subsection{Comparison with a Modern GPU}

\wqg{We further evaluate \system against a more recent GPU platform, NVIDIA H100, from a normalized-efficiency perspective. Ascend 910B provides 280 TFLOPS of peak compute throughput and 1.8 TB/s memory bandwidth, whereas H100 provides 1600 TFLOPS and 2.0 TB/s. Given this large gap in raw compute capability, a direct throughput comparison would be less fair. We therefore normalize performance by peak compute throughput (TFLOPS) and compare the resulting efficiency across platforms. As shown in Fig.~\ref{fig:compare-H100}, \system on Ascend 910B achieves higher normalized efficiency than both cuSPARSE and DTC-SpMM on H100 for most datasets, with average improvements of $1.64\times$ and $1.55\times$, respectively. This result suggests that, despite the much stronger absolute compute capability of H100, \system still delivers competitive sparse-computation efficiency.}

\subsection{\zyf{Portability to Another NPU Backend}}

\begin{figure}[t]
  \centering
  \includegraphics[width=0.7\linewidth]{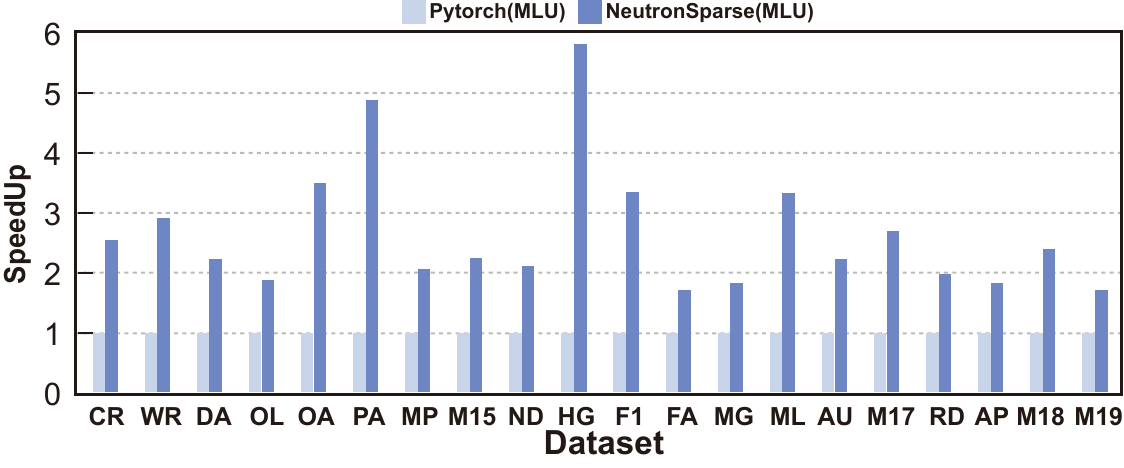}
  \vspace{-0.1in}
  \caption{{\zyf{SpMM performance on Cambricon MLU across all 20 datasets.}}}
  \label{fig:compare-MLU}
  \vspace{-0.1in}
\end{figure}

To further evaluate cross-device portability, we port \system to Cambricon MLU and compare it against the SpMM implementation provided by PyTorch on the same platform.
The port preserves the core ideas of \system by adapting heterogeneous workload 
partitioning and tile-aware scheduling to the execution primitives and software stack of MLU. 
Compared with Ascend 910B, Cambricon MLU also adopts a heterogeneous design with separate vector and matrix processing engines, but differs in its execution abstraction and memory hierarchy, where tensor operations rely more on compiler-managed tiling and shared on-chip buffers, leading to tighter coupling between engines and requiring additional backend-specific memory management and scheduling support for coordinated execution.
As shown in Fig.~\ref{fig:compare-MLU}, we evaluate all 20 datasets and observe consistent improvements across the full set: \system achieves speedups ranging from about $1.71\times$ to $5.83\times$, with the largest gain on \texttt{HG} and clear improvements on most datasets. These results suggest that, although backend-specific engineering is still required, the main design principles of \system are not limited to Ascend and can also deliver substantial benefits on another NPU family.

\section{Related Work}
\label{sec8}

\Paragraph{SpMM Framework.}  
Many works have sought to accelerate SpMM workloads arising in GNNs \cite{yanRTGNNAcceleratingSparse2024,wangTCGNNBridgingSparse2021,sarkarFlowGNNDataflowArchitecture2022, merkelCanGraphReordering2024, yanMulticoatedFoldedGraph2023}, graph analytics \cite{zouSurveyKnowledgeGraph2020,yanMulticoatedFoldedGraph2023,merkelCanGraphReordering2024}, and sparse attentions \cite{liEfficientQuantizedSparse2022,roy2021attention}. Most GPU-based approaches utilizing a single compute unit within an SM (CUDA cores or Tensor Cores) \cite{galeSparseGPUKernels2020,kurtEfficientTiledSparse2020,hongAdaptiveSparseTiling2019,xuVectorSparseRepresentation2015,huangGESpMMGeneralPurposeSparse2020,yangDesignPrinciplesSparse2018,zhuSparseTensorCore2019,fanDTCSpMMBridgingGap2024,liHCSpMMAcceleratingSparse2024}. For example, ASpT \cite{hongAdaptiveSparseTiling2019} employs column-block reordering to form dense tiles and improve data locality; RoDe \cite{pangRoDe2024} applies row decomposition to normalize long and short rows for better load balance; and Acc-SpMM \cite{zhaoAccSpMMAcceleratingGeneralpurpose2025} leverages hardware–software co-design to optimize the \texttt{wmma} API for sparse workloads.  
Recent studies have explored heterogeneous cooperation between CUDA cores and Tensor Cores. For example, HC-SpMM \cite{liHCSpMMAcceleratingSparse2024} and Libra \cite{shi2025libra} investigate hybrid execution modes by leveraging both CUDA cores and Tensor Cores on GPUs. 

\Paragraph{Research on Ascend NPUs.}  
Research on Ascend NPUs is gradually gaining traction in the systems and architecture community\cite{zhuPerformanceEvaluationMindSpore2023,ascendwork1, Ascend_HPCA_2021}. MindSporeGL \cite{zhuPerformanceEvaluationMindSpore2023} provides a GNN training framework on MindSpore; NeutronAscend \cite{AIXIN2025neutronascend} proposes a heterogeneous scheduling framework that dispatches light and heavy workloads across units; and Zhou \emph{et al.} present an automatic operator compilation technique alongside a componentized Roofline model for operator performance analysis~\cite{ascendwork1,ascendwork2}.
XY-Serve \cite{song2026xy} proposed an end-to-end LLM service system that efficiently adapts to hardware by decomposing, rearranging, and abstracting dynamic workloads.




\begin{table}[t]
  \centering
  \caption{Common architectural features leveraged by NeutronSparse across NPU families.}
  \vspace{-0.15in}
  \renewcommand{\arraystretch}{0.95}
  \setlength{\tabcolsep}{4pt}
  \scriptsize
  \begin{tabular}{l c c}
    \toprule
    \textbf{NPU} 
    & \textbf{Heterogeneous Compute Engines} 
    & \textbf{Tile-Based Execution} \\
    \midrule
    Ascend 910B \cite{Ascend_HPCA_2021} 
      & \checkmark (AIC + AIV) 
      & \checkmark \\

    TPU v4 \cite{TPU_ISCA_2017}         
      & \checkmark (MXU (Systolic Array) + vector/scalar units) 
      & \checkmark \\

    Gaudi2 \cite{Gaudi_Habana_2020}     
      & \checkmark (MME (Matrix Math Engine)  + vector cores) 
      & \checkmark \\

    MLU370 \cite{MLU_MICRO_2021}        
      & \checkmark (MMU (Matrix Multiply Unit)  + vector units) 
      & \checkmark \\
    \bottomrule
  \end{tabular}
  \vspace{-0.10in}
  \label{tab:neuronsparse_generalization}
\end{table}

\section{Future Work}

\Paragraph{Extension to Other NPUs.}
\wqg{Although NeutronSparse is implemented on Ascend, its high-level design is not fully Ascend-specific.}
It is grounded in two architectural properties that are increasingly common across modern NPUs:
\textbf{(1) matrix-centric heterogeneous engines} and \textbf{(2) tile-based execution}.
As summarized in Table~\ref{tab:neuronsparse_generalization}, mainstream NPUs (e.g., TPU \cite{TPU_ISCA_2017}, Gaudi \cite{Gaudi_Habana_2020}, and MLU \cite{MLU_MICRO_2021}) integrate dedicated matrix engines together with vector/scalar units, and expose fixed-shape tiles as the basic execution granularity.
Under sparse workloads, this organization makes performance hinge on (i) coordinating heterogeneous engines to avoid idle resources and (ii) scheduling work at tile granularity to mitigate tile-level inefficiency caused by irregular sparsity.
\wqg{Accordingly, the high-level ideas of heterogeneous workload partitioning and tile-aware scheduling can generalize to other NPUs. In contrast, AIV-AIC coordinated pipelining is more Ascend-specific, since it depends on the decoupled AIC/AIV design. On other NPUs, such as Cambricon MLU and Google TPU, heterogeneous engines may still share and contend for local buffers and on-chip memory resources, so coordinated execution requires additional backend-specific memory management and scheduling support.} 

\Paragraph{\wqg{Learning-based Initialization.}}
\wqg{Our current design combines a lightweight initial split with online migration, but the optimal AIV/AIC ratio can still be affected by dynamic factors such as irregular sparsity, memory interference, and input-dependent behavior. 
A more accurate initial ratio could further reduce the warm-up period, especially for short-running workloads where online adjustment has less time to amortize.
A promising direction for future work is learning-based initialization, which predicts a better initial split from matrix features and runtime signals.}

\section{Conclusion}
\label{sec9}

This paper shows that efficient SpMM on Ascend NPUs requires NPU-specific optimization rather than directly porting GPU designs. We develop \system, which addresses two bottlenecks: (i) balancing heterogeneous engines (AIC/AIV) via architecture-aware partitioning and online tile migration, and (ii) mitigating tile-level sparsity inefficiency via locality-aware tile orchestrating (global-local reordering and hierarchical tile reuse). Experiments on Ascend 910B demonstrate substantial speedups over NPU baselines and competitive performance against state-of-the-art GPU libraries.



\begin{acks}
 This work was supported by the National Natural Science Foundation of China (62461146205, U2241212, 625B2041), and the Distinguished Youth Foundation of Liaoning Province (2024021148-JH3/501).
\end{acks}

\bibliographystyle{ACM-Reference-Format}
\bibliography{sample-base}

@String{Computing = "Computing" }

@String{Computer = "{IEEE} Computer" }

@article{ahmadExploringDataLayout2024,
  author  = {Ahmad, Khalid and Cecka, Cris and Garland, Michael and Hall, Mary},
  title   = {Exploring Data Layout for Sparse Tensor Times Dense Matrix on GPUs},
  journal = {ACM Transactions on Architecture and Code Optimization},
  volume  = {21},
  number  = {1},
  pages   = {1--20},
  year    = {2024}
}

@misc{dharAscendCCConfidentialComputing2024,
  author       = {Dhar, Aritra and Thorens, Clément and Lazier, Lara Magdalena and Cavigelli, Lukas},
  title        = {Ascend-CC: Confidential Computing on Heterogeneous NPU for Emerging Generative AI Workloads},
  howpublished = {\url{https://arxiv.org/abs/2407.11888}},
  year         = {2024}
}

@article{duffSurveySparseMatrix1977,
  author  = {Duff, I.~S.},
  title   = {A Survey of Sparse Matrix Research},
  journal = {Proceedings of the IEEE},
  volume  = {65},
  number  = {4},
  pages   = {500--535},
  year    = {1977}
}

@inproceedings{fanDTCSpMMBridgingGap2024,
  author    = {Fan, Ruibo and Wang, Wei and Chu, Xiaowen},
  title     = {{DTC-SpMM}: Bridging the Gap in Accelerating General Sparse Matrix Multiplication with Tensor Cores},
  booktitle = {Proceedings of the 29th ACM International Conference on Architectural Support for Programming Languages and Operating Systems (ASPLOS), Volume~3},
  pages     = {253--267},
  year      = {2024}
}

@misc{galeSparseGPUKernels2020,
  author       = {Gale, Trevor and Zaharia, Matei and Young, Cliff and Elsen, Erich},
  title        = {Sparse GPU Kernels for Deep Learning},
  howpublished = {\url{https://arxiv.org/abs/2006.10901}},
  year         = {2020}
}

@inproceedings{heoNeuPIMsNPUPIMHeterogeneous2024,
  author    = {Heo, Guseul and Lee, Sangyeop and Cho, Jaehong and Choi, Hyunmin and Lee, Sanghyeon and Ham, Hyungkyu and Kim, Gwangsun and Mahajan, Divya and Park, Jongse},
  title     = {{NeuPIMs}: {NPU–PIM} Heterogeneous Acceleration for Batched LLM Inferencing},
  booktitle = {Proceedings of the 29th ACM International Conference on Architectural Support for Programming Languages and Operating Systems (ASPLOS), Volume~3},
  pages     = {722--737},
  year      = {2024}
}

@inproceedings{hoImprovingGPUThroughput2022,
  author    = {Ho, Khoa and Zhao, Hui and Jog, Adwait and Mohanty, Saraju},
  title     = {Improving GPU Throughput through Parallel Execution Using Tensor Cores and CUDA Cores},
  booktitle = {IEEE Computer Society Annual Symposium on VLSI (ISVLSI)},
  pages     = {223--228},
  year      = {2022}
}

@inproceedings{hongAdaptiveSparseTiling2019,
  author    = {Hong, Changwan and Sukumaran-Rajam, Aravind and Nisa, Israt and Singh, Kunal and Sadayappan, P.},
  title     = {Adaptive Sparse Tiling for Sparse Matrix Multiplication},
  booktitle = {Proceedings of the 24th ACM SIGPLAN Symposium on Principles and Practice of Parallel Programming (PPoPP)},
  pages     = {300--314},
  year      = {2019}
}

@inproceedings{huangGESpMMGeneralPurposeSparse2020,
  author    = {Huang, Guyue and Dai, Guohao and Wang, Yu and Yang, Huazhong},
  title     = {{GE-SpMM}: General-Purpose Sparse Matrix–Matrix Multiplication on GPUs for Graph Neural Networks},
  booktitle = {SC20: International Conference for High Performance Computing, Networking, Storage and Analysis},
  pages     = {1--12},
  year      = {2020}
}

@inproceedings{kurtEfficientTiledSparse2020,
  author    = {Kurt, Sureyya Emre and Sukumaran-Rajam, Aravind and Rastello, Fabrice and Sadayyapan, P.},
  title     = {Efficient Tiled Sparse Matrix Multiplication through Matrix Signatures},
  booktitle = {SC20: International Conference for High Performance Computing, Networking, Storage and Analysis},
  pages     = {1--14},
  year      = {2020}
}

@inproceedings{liEfficientQuantizedSparse2022,
  author    = {Li, Shigang and Osawa, Kazuki and Hoefler, Torsten},
  title     = {Efficient Quantized Sparse Matrix Operations on Tensor Cores},
  booktitle = {SC22: International Conference for High Performance Computing, Networking, Storage and Analysis},
  pages     = {1--15},
  year      = {2022}
}

@inproceedings{liHCSpMMAcceleratingSparse2024,
  author       = {Zhonggen Li and
                  Xiangyu Ke and
                  Yifan Zhu and
                  Yunjun Gao and
                  Yaofeng Tu},
  title        = {HC-SpMM: Accelerating Sparse Matrix-Matrix Multiplication for Graphs
                  with Hybrid {GPU} Cores},
  booktitle    = {41st {IEEE} International Conference on Data Engineering, {ICDE} 2025,
                  Hong Kong, May 19-23, 2025},
  pages        = {501--514},
  publisher    = {{IEEE}},
  year         = {2025}
}

@inproceedings{luoSparmSparseMatrix2024,
  author    = {Luo, Shengbai and Wang, Bo and Shi, Yihao and Zhang, Xueyi and Xue, Qingshan and Ma, Sheng},
  title     = {{Sparm}: A Sparse Matrix Multiplication Accelerator Supporting Multiple Dataflows},
  booktitle = {IEEE 35th International Conference on Application-specific Systems, Architectures and Processors (ASAP)},
  pages     = {122--130},
  year      = {2024}
}

@article{merkelCanGraphReordering2024,
  author  = {Merkel, Nikolai and Toussing, Pierre and Mayer, Ruben and Jacobsen, Hans-Arno},
  title   = {Can Graph Reordering Speed Up Graph Neural Network Training? An Experimental Study},
  journal = {Proceedings of the VLDB Endowment},
  volume  = {18},
  number  = {2},
  pages   = {293--307},
  year    = {2024}
}

@misc{sarkarFlowGNNDataflowArchitecture2022,
  author       = {Sarkar, Rishov and Abi-Karam, Stefan and He, Yuqi and Sathidevi, Lakshmi and Hao, Cong},
  title        = {{FlowGNN}: A Dataflow Architecture for Real-Time Workload-Agnostic Graph Neural Network Inference},
  howpublished = {\url{https://arxiv.org/abs/2204.13103}},
  year         = {2022}
}

@inproceedings{srivastavaMatRaptorSparseSparseMatrix2020,
  author    = {Srivastava, Nitish and Jin, Hanchen and Liu, Jie and Albonesi, David and Zhang, Zhiru},
  title     = {{MatRaptor}: A Sparse–Sparse Matrix Multiplication Accelerator Based on Row-Wise Product},
  booktitle = {53rd Annual IEEE/ACM International Symposium on Microarchitecture (MICRO)},
  pages     = {766--780},
  year      = {2020}
}

@inproceedings{wangSwSpTRSVFastSparse2018,
  author    = {Wang, Xinliang and Liu, Weifeng and Xue, Wei and Wu, Li},
  title     = {{swSpTRSV}: A Fast Sparse Triangular Solve with Sparse Level Tile Layout on Sunway Architectures},
  booktitle = {Proceedings of the 23rd ACM SIGPLAN Symposium on Principles and Practice of Parallel Programming (PPoPP)},
  pages     = {338--353},
  year      = {2018}
}

@article{wangTCGNNBridgingSparse2021,
  author  = {Wang, Yuke and Feng, Boyuan and Wang, Zheng and Huang, Guyue and Ding, Yufei},
  title   = {{TC-GNN}: Bridging Sparse GNN Computation and Dense Tensor Cores on GPUs},
  journal = {IEEE Transactions on Parallel and Distributed Systems},
  year    = {2021},
}

@article{xiePatternBasedSpGEMMLibrary2022,
  author  = {Xie, Zhen and Tan, Guangming and Liu, Weifeng and Sun, Ninghui},
  title   = {A Pattern-Based SpGEMM Library for Multi-Core and Many-Core Architectures},
  journal = {IEEE Transactions on Parallel and Distributed Systems},
  volume  = {33},
  number  = {1},
  pages   = {159--175},
  year    = {2022}
}

@article{xuVectorSparseRepresentation2015,
  author  = {Xu, Yi and Yu, Licheng and Xu, Hongteng and Zhang, Hao and Nguyen, Truong},
  title   = {Vector Sparse Representation of Color Image Using Quaternion Matrix Analysis},
  journal = {IEEE Transactions on Image Processing},
  volume  = {24},
  number  = {4},
  pages   = {1315--1329},
  year    = {2015}
}

@misc{yangDesignPrinciplesSparse2018,
  author       = {Yang, Carl and Buluc, Aydin and Owens, John~D.},
  title        = {Design Principles for Sparse Matrix Multiplication on the GPU},
  howpublished = {\url{https://arxiv.org/abs/1803.08601}},
  year         = {2018}
}

@misc{yanMulticoatedFoldedGraph2023,
  author       = {Yan, Jiale and Ito, Hiroaki and Garcia-Arias, Angel Lopez and Okoshi, Yasuyuki and Otsuka, Hikari and Kawamura, Kazushi and Chu, Thiem Van and Motomura, Masato},
  title        = {Multicoated and Folded Graph Neural Networks with Strong Lottery Tickets},
  howpublished = {Manuscript},
  year         = {2023}
}

@article{yanRTGNNAcceleratingSparse2024,
  author  = {Yan, Jianrong and Jiang, Wenbin and He, Dongao and Wen, Suyang and Li, Yang and Jin, Hai and Shao, Zhiyuan},
  title   = {{RT-GNN}: Accelerating Sparse Graph Neural Networks by Tensor–CUDA Kernel Fusion},
  journal = {ACM Transactions on Architecture and Code Optimization},
  pages   = {3702001},
  year    = {2024}
}

@inproceedings{pangRoDe2024,
  author    = {Pang, Meng and Fei, Xiang and Qu, Peng and Zhang, Youhui and Li, Zhaolin},
  title     = {{RoDe}: A Row Decomposition–Based Approach for Sparse Matrix Multiplication on GPUs},
  booktitle = {Proceedings of the 29th ACM SIGPLAN Annual Symposium on Principles and Practice of Parallel Programming (PPoPP)},
  pages     = {377--389},
  year      = {2024}
}

@article{zachariadisAcceleratingSparseMatrixMatrix2020,
  author  = {Zachariadis, Orestis and Satpute, Nitin and Gómez-Luna, Juan and Olivares, Joaquín},
  title   = {Accelerating Sparse Matrix–Matrix Multiplication with GPU Tensor Cores},
  journal = {Computers \& Electrical Engineering},
  volume  = {88},
  pages   = {106848},
  year    = {2020}
}

@inproceedings{zhaoAccSpMMAcceleratingGeneralpurpose2025,
  author    = {Zhao, Haisha and Li, San and Wang, Jiaheng and Zhou, Chunbao and Wang, Jue and Xin, Zhikuang and Li, Shunde and Liang, Zhiqiang and Pan, Zhijie and Liu, Fang and Zeng, Yan and Wang, Yangang and Chi, Xuebin},
  title     = {Acc-SpMM: Accelerating General-Purpose Sparse Matrix–Matrix Multiplication with GPU Tensor Cores},
  booktitle = {Proceedings of the 30th ACM SIGPLAN Annual Symposium on Principles and Practice of Parallel Programming (PPoPP)},
  pages     = {326--338},
  year      = {2025}
}

@inproceedings{zhaoExploitingIntraSMParallelism2021,
  author    = {Zhao, Han and Cui, Weihao and Chen, Quan and Zhao, Jieru and Leng, Jingwen and Guo, Minyi},
  title     = {Exploiting Intra-SM Parallelism in GPUs via Persistent and Elastic Blocks},
  booktitle = {IEEE International Conference on Computer Design (ICCD)},
  pages     = {290--298},
  year      = {2021}
}

@inproceedings{zhuPerformanceEvaluationMindSpore2023,
  author    = {Zhu, Zeling and Wang, Bangchuan and Yang, Chuying and Zhu, Rui and Zhou, Mingyao and Zheng, Nenggan},
  title     = {Performance Evaluation of MindSpore and PyTorch Based on Ascend NPU},
  booktitle = {IEEE International Conference on Parallel and Distributed Systems (ICPADS)},
  pages     = {1826--1832},
  year      = {2023}
}

@misc{nvidia2023cusparse,
  author       = {{NVIDIA}},
  title        = {CUDA Sparse Matrix Library},
  howpublished = {\url{https://developer.nvidia.com/cusparse}},
  year         = {2023},
  note         = {Accessed 2025-05-09}
}

@misc{nvidia2023cublas,
  author       = {{NVIDIA}},
  title        = {Dense Linear Algebra on GPUs},
  howpublished = {\url{https://developer.nvidia.com/cublas}},
  year         = {2023},
  note         = {Accessed 2025-05-09}
}

@inproceedings{zhuSparseTensorCore2019,
  author    = {Zhu, Maohua and Zhang, Tao and Gu, Zhenyu and Xie, Yuan},
  title     = {Sparse Tensor Core: Algorithm and Hardware Co-Design for Vector-Wise Sparse Neural Networks on Modern GPUs},
  booktitle = {52nd Annual IEEE/ACM International Symposium on Microarchitecture (MICRO)},
  pages     = {359--371},
  year      = {2019}
}

@article{kosurveyrecommendationsystems2022,
  author  = {Ko, Hyeyoung and Lee, Suyeon and Park, Yoonseo and Choi, Anna},
  title   = {A Survey of Recommendation Systems: Recommendation Models, Techniques, and Application Fields},
  journal = {Electronics},
  volume  = {11},
  number  = {1},
  pages   = {141},
  year    = {2022}
}

@article{waltersMolecularPropertyPrediction2020,
  author  = {Walters, W. Patrick and Barzilay, Regina},
  title   = {Applications of Deep Learning in Molecule Generation and Molecular Property Prediction},
  journal = {Accounts of Chemical Research},
  volume  = {54},
  number  = {2},
  pages   = {263--270},
  year    = {2020}
}

@inproceedings{zouSurveyKnowledgeGraph2020,
  author    = {Zou, Xiaohan},
  title     = {A Survey on Application of Knowledge Graph},
  booktitle = {Journal of Physics: Conference Series},
  volume    = {1487},
  number    = {1},
  pages     = {012016},
  year      = {2020}
}

@inproceedings{MLU_MICRO_2021,
  author       = {Shaoli Liu and
                  Zidong Du and
                  Jinhua Tao and
                  Dong Han and
                  Tao Luo and
                  Yuan Xie and
                  Yunji Chen and
                  Tianshi Chen},
  title        = {Cambricon: An Instruction Set Architecture for Neural Networks},
  booktitle    = {43rd {ACM/IEEE} Annual International Symposium on Computer Architecture,
                  {ISCA} 2016, Seoul, South Korea, June 18-22, 2016},
  pages        = {393--405},
  publisher    = {{IEEE} Computer Society},
  year         = {2016}
}

@article{Gaudi_Habana_2020,
  author       = {Eitan Medina and
                  Eran Dagan},
  title        = {Habana Labs Purpose-Built {AI} Inference and Training Processor Architectures:
                  Scaling {AI} Training Systems Using Standard Ethernet With Gaudi Processor},
  journal      = {{IEEE} Micro},
  volume       = {40},
  number       = {2},
  pages        = {17--24},
  year         = {2020}
}

@inproceedings{TPU_ISCA_2017,
  author       = {Norman P. Jouppi and
                  Cliff Young and
                  Nishant Patil and
                  David A. Patterson and
                  Gaurav Agrawal and
                  Raminder Bajwa and
                  Sarah Bates and
                  Suresh Bhatia and
                  Nan Boden and
                  Al Borchers and
                  Rick Boyle and
                  Pierre{-}luc Cantin and
                  Clifford Chao and
                  Chris Clark and
                  Jeremy Coriell and
                  Mike Daley and
                  Matt Dau and
                  Jeffrey Dean and
                  Ben Gelb and
                  Tara Vazir Ghaemmaghami and
                  Rajendra Gottipati and
                  William Gulland and
                  Robert Hagmann and
                  C. Richard Ho and
                  Doug Hogberg and
                  John Hu and
                  Robert Hundt and
                  Dan Hurt and
                  Julian Ibarz and
                  Aaron Jaffey and
                  Alek Jaworski and
                  Alexander Kaplan and
                  Harshit Khaitan and
                  Daniel Killebrew and
                  Andy Koch and
                  Naveen Kumar and
                  Steve Lacy and
                  James Laudon and
                  James Law and
                  Diemthu Le and
                  Chris Leary and
                  Zhuyuan Liu and
                  Kyle Lucke and
                  Alan Lundin and
                  Gordon MacKean and
                  Adriana Maggiore and
                  Maire Mahony and
                  Kieran Miller and
                  Rahul Nagarajan and
                  Ravi Narayanaswami and
                  Ray Ni and
                  Kathy Nix and
                  Thomas Norrie and
                  Mark Omernick and
                  Narayana Penukonda and
                  Andy Phelps and
                  Jonathan Ross and
                  Matt Ross and
                  Amir Salek and
                  Emad Samadiani and
                  Chris Severn and
                  Gregory Sizikov and
                  Matthew Snelham and
                  Jed Souter and
                  Dan Steinberg and
                  Andy Swing and
                  Mercedes Tan and
                  Gregory Thorson and
                  Bo Tian and
                  Horia Toma and
                  Erick Tuttle and
                  Vijay Vasudevan and
                  Richard Walter and
                  Walter Wang and
                  Eric Wilcox and
                  Doe Hyun Yoon},
  title        = {In-Datacenter Performance Analysis of a Tensor Processing Unit},
  booktitle    = {Proceedings of the 44th Annual International Symposium on Computer
                  Architecture, {ISCA} 2017, Toronto, ON, Canada, June 24-28, 2017},
  pages        = {1--12},
  publisher    = {{ACM}},
  year         = {2017}
}

@inproceedings{ascendwork1,
  author       = {Yuhang Zhou and
                  Zibo Wang and
                  Zhibin Wang and
                  Ruyi Zhang and
                  Chen Tian and
                  Xiaoliang Wang and
                  Wanchun Dou and
                  Guihai Chen and
                  Bingqiang Wang and
                  Yonghong Tian and
                  Yan Zhang and
                  Hui Wang and
                  Fuchun Wei and
                  Boquan Sun and
                  Jingyi Zhang and
                  Bin She and
                  Teng Su and
                  Yifan Yao and
                  Chunsheng Li and
                  Ziyang Zhang and
                  Yaoyuan Wang and
                  Bin Zhou and
                  Guyue Liu},
  editor       = {Deniz Altinb{\"{u}}ken and
                  Ryan Stutsman},
  title        = {Accelerating Model Training on Ascend Chips: An Industrial System
                  for Profiling, Analysis and Optimization},
  booktitle    = {Proceedings of the 2025 {USENIX} Annual Technical Conference, {USENIX}
                  {ATC} 2025, Boston, MA, USA, July 7-9, 2025},
  pages        = {1387--1408},
  publisher    = {{USENIX} Association},
  year         = {2025}
}

@inproceedings{ascendwork2,
  author       = {Yuhang Zhou and
                  Zhibin Wang and
                  Guyue Liu and
                  Shipeng Li and
                  Xi Lin and
                  Zibo Wang and
                  Yongzhong Wang and
                  Fuchun Wei and
                  Jingyi Zhang and
                  Zhiheng Hu and
                  Yanlin Liu and
                  Chunsheng Li and
                  Ziyang Zhang and
                  Yaoyuan Wang and
                  Bin Zhou and
                  Wanchun Dou and
                  Guihai Chen and
                  Chen Tian},
  editor       = {Lieven Eeckhout and
                  Georgios Smaragdakis and
                  Katai Liang and
                  Adrian Sampson and
                  Martha A. Kim and
                  Christopher J. Rossbach},
  title        = {Squeezing Operator Performance Potential for the Ascend Architecture},
  booktitle    = {Proceedings of the 30th {ACM} International Conference on Architectural
                  Support for Programming Languages and Operating Systems, Volume 2,
                  {ASPLOS} 2025, Rotterdam, Netherlands, 30 March 2025 - 3 April 2025},
  pages        = {1156--1171},
  publisher    = {{ACM}},
  year         = {2025}
}

@article{AIXIN2025neutronascend,
  title={NeutronAscend: Optimizing GNN Training with Ascend AI
Processors},
  author={Xin Ai and
          Bing Zhang and
          Qiange Wang and
          Yanfeng Zhang and
          Hao Yuan and
          Shufeng Gong and
          Ge Yu},
  journal={ACM Transactions on Architecture and Code Optimization},
  year={2025}
}

@inproceedings{Ascend_HPCA_2021,
  author       = {Heng Liao and
                  Jiajin Tu and
                  Jing Xia and
                  Hu Liu and
                  Xiping Zhou and
                  Honghui Yuan and
                  Yuxing Hu},
  title        = {Ascend: a Scalable and Unified Architecture for Ubiquitous Deep Neural
                  Network Computing : Industry Track Paper},
  booktitle    = {{IEEE} International Symposium on High-Performance Computer Architecture,
                  {HPCA} 2021, Seoul, South Korea, February 27 - March 3, 2021},
  pages        = {789--801},
  publisher    = {{IEEE}},
  year         = {2021}
}

@inproceedings{DaVinci_HCS_2019,
  author       = {Heng Liao and
                  Jiajin Tu and
                  Jing Xia and
                  Xiping Zhou},
  title        = {DaVinci: {A} Scalable Architecture for Neural Network Computing},
  booktitle    = {2019 {IEEE} Hot Chips 31 Symposium (HCS), Cupertino, CA, USA, August
                  18-20, 2019},
  pages        = {1--44},
  publisher    = {{IEEE}},
  year         = {2019}
}

@inproceedings{arai2016rabbit,
  title={Rabbit order: Just-in-time parallel reordering for fast graph analysis},
  author={Arai, Junya and Shiokawa, Hiroaki and Yamamuro, Takeshi and Onizuka, Makoto and Iwamura, Sotetsu},
  booktitle={2016 IEEE International Parallel and Distributed Processing Symposium (IPDPS)},
  pages={22--31},
  year={2016}
}

@article{davis2011suitesparse,
  title={The University of Florida sparse matrix collection},
  author={Davis, Timothy A and Hu, Yifan},
  journal={ACM Transactions on Mathematical Software (TOMS)},
  volume={38},
  number={1},
  pages={1--25},
  year={2011}
}

@article{mccallum2000cora,
  title={Automating the construction of internet portals with machine learning},
  author={McCallum, Andrew Kachites and Nigam, Kamal and Rennie, Jason and Seymore, Kristie},
  journal={Information Retrieval},
  volume={3},
  number={2},
  pages={127--163},
  year={2000}
}

@article{hu2020ogbn-arxiv,
  title={Open graph benchmark: Datasets for machine learning on graphs},
  author={Hu, Weihua and Fey, Matthias and Zitnik, Marinka and Dong, Yuxiao and Ren, Hongyu and Liu, Bowen and Catasta, Michele and Leskovec, Jure},
  journal={Advances in neural information processing systems},
  volume={33},
  pages={22118--22133},
  year={2020}
}

@article{hamilton2017reddit,
  title={Inductive representation learning on large graphs},
  author={Hamilton, Will and Ying, Zhitao and Leskovec, Jure},
  journal={Advances in neural information processing systems},
  volume={30},
  year={2017}
}

@inproceedings{mcauley2015amazon,
  title={Image-based recommendations on styles and substitutes},
  author={McAuley, Julian and Targett, Christopher and Shi, Qinfeng and Van Den Hengel, Anton},
  booktitle={Proceedings of the 38th international ACM SIGIR conference on research and development in information retrieval},
  pages={43--52},
  year={2015}
}

@article{roy2021attention,
  title={Efficient content-based sparse attention with routing transformers},
  author={Roy, Aurko and Saffar, Mohammad and Vaswani, Ashish and Grangier, David},
  journal={Transactions of the Association for Computational Linguistics},
  volume={9},
  pages={53--68},
  year={2021}
}

@article{shi2025libra,
  title={Libra: Synergizing CUDA and Tensor Cores for High-Performance Sparse Matrix Multiplication},
  author={Shi, Jinliang and Li, Shigang and Xu, Youxuan and Wang, Xueying and Fu, Rongtian and Ma, Zhi and Wu, Tong},
  journal={arXiv preprint arXiv:2506.22714},
  year={2025}
}

@inproceedings{shi2025flashsparse,
  title={Flashsparse: Minimizing computation redundancy for fast sparse matrix multiplications on tensor cores},
  author={Shi, Jinliang and Li, Shigang and Xu, Youxuan and Fu, Rongtian and Wang, Xueying and Wu, Tong},
  booktitle={Proceedings of the 30th ACM SIGPLAN Annual Symposium on Principles and Practice of Parallel Programming},
  pages={312--325},
  year={2025}
}

@inproceedings{song2026xy,
  title={XY-Serve: End-to-End Versatile Production Serving for Dynamic LLM Workloads},
  author={Song, Mingcong and Tang, Xinru and Hou, Fengfan and Li, Jing and Wei, Wei and Ma, Yipeng and Xiao, Runqiu and Si, Hongjie and Jiang, Dingcheng and Yin, Shouyi and others},
  booktitle={Proceedings of the 31st ACM International Conference on Architectural Support for Programming Languages and Operating Systems, Volume 1},
  pages={314--329},
  year={2026}
}

@inproceedings{neutron_sigmod_2022,
  author    = {Qiange Wang and
               Yanfeng Zhang and
               Hao Wang and
               Chaoyi Chen and
               Xiaodong Zhang and
               Ge Yu},
  title     = {NeutronStar: Distributed {GNN} Training with Hybrid Dependency Management},
  booktitle = {International Conference on Management of Data, Philadelphia, SIGMOD'22,
               PA, USA},
  pages     = {1301--1315},
  publisher = {{ACM}},
  year      = {2022}
}

@article{G3_SIGMOD_2023,
  author       = {Xinchen Wan and
                  Kaiqiang Xu and
                  Xudong Liao and
                  Yilun Jin and
                  Kai Chen and
                  Xin Jin},
  title        = {Scalable and Efficient Full-Graph {GNN} Training for Large Graphs},
  journal      = {Proc. {ACM} Manag. Data},
  volume       = {1},
  number       = {2},
  pages        = {143:1--143:23},
  year         = {2023}
}

@article{hongtu_SIGMOD_2024,
  author       = {Qiange Wang and
                  Yao Chen and
                  Weng{-}Fai Wong and
                  Bingsheng He},
  title        = {HongTu: Scalable Full-Graph {GNN} Training on Multiple GPUs},
  journal      = {Proc. {ACM} Manag. Data},
  volume       = {1},
  number       = {4},
  pages        = {246:1--246:27},
  year         = {2023}
}

@article{SANCUS_VLDB_2022,
  author    = {Jingshu Peng and
               Zhao Chen and
               Yingxia Shao and
               Yanyan Shen and
               Lei Chen and
               Jiannong Cao},
  title     = {{SANCUS:} Staleness-Aware Communication-Avoiding Full-Graph Decentralized
               Training in Large-Scale Graph Neural Networks},
  journal   = {Proc. {VLDB} Endow.},
  volume    = {15},
  number    = {9},
  pages     = {1937--1950},
  year      = {2022}
}

@article{NeutronOrch2024ai,
  author       = {Xin Ai and
                  Qiange Wang and
                  Chunyu Cao and
                  Yanfeng Zhang and
                  Chaoyi Chen and
                  Hao Yuan and
                  Yu Gu and
                  Ge Yu},
  title        = {NeutronOrch: Rethinking Sample-based {GNN} Training under {CPU-GPU}
                  Heterogeneous Environments},
  journal      = {Proc. {VLDB} Endow.},
  volume       = {17},
  number       = {8},
  pages        = {1995--2008},
  year         = {2024},
}

@article{aligraph_vldb_2019,
  author       = {Rong Zhu and
                  Kun Zhao and
                  Hongxia Yang and
                  Wei Lin and
                  Chang Zhou and
                  Baole Ai and
                  Yong Li and
                  Jingren Zhou},
  title        = {AliGraph: {A} Comprehensive Graph Neural Network Platform},
  journal      = {Proc. {VLDB} Endow.},
  volume       = {12},
  number       = {12},
  pages        = {2094--2105},
  year         = {2019}
}

@inproceedings{moustafa2023accelerating,
  title={Accelerating sparse matrix-matrix multiplication with the Ascend AI core},
  author={Moustafa, Salli},
  booktitle={Proc. Workshop Accelerated Mach. Learn},
  pages={33--41},
  year={2023}
}

@article{neutrontp_vldb24,
author = {Ai, Xin and Yuan, Hao and Ling, Zeyu and Wang, Qiange and Zhang, Yanfeng and Fu, Zhenbo and Chen, Chaoyi and Gu, Yu and Yu, Ge},
title = {NeutronTP: Load-Balanced Distributed Full-Graph GNN Training with Tensor Parallelism},
journal = {Proceedings of the VLDB Endowment},
volume = {18},
number = {2},
pages = {173–186},
year = {2024},
}

\end{document}